\documentclass[prd,showpacs,preprintnumbers,amsmath,amssymb]{revtex4}
\usepackage{cancel}
\usepackage{booktabs}
\usepackage{multirow}
\usepackage{slashed}
\usepackage{graphicx}
\usepackage[colorlinks,
            citecolor=blue,
            anchorcolor=red,
            menucolor=red,
            linkcolor=red,
            filecolor=red,
            runcolor=red,
            urlcolor=blue,
            frenchlinks=red]{hyperref}

\begin{document}
\title{Hidden-charm pentaquarks and their hidden-bottom and $B_c$-like partner states}
\author{Jing Wu$^1$}
\author{Yan-Rui Liu$^{1,2}$}
\email{yrliu@sdu.edu.cn} \affiliation{ $^1$School of Physics and Key
Laboratory of Particle Physics and Particle Irradiation (MOE), Shandong University, Jinan 250100, China\\
$^2$Key Laboratory of Theoretical Physics, Institute of Theoretical
Physics, CAS, Beijing 100190, China}
\author{Kan Chen$^{3,4}$}
\author{Xiang Liu$^{3,4}$}
\email{xiangliu@lzu.edu.cn} \affiliation{
$^3$School of Physical Science and Technology, Lanzhou University, Lanzhou 730000, China\\
$^4$Research Center for Hadron and CSR Physics, Lanzhou University
and Institute of Modern Physics of CAS, Lanzhou 730000, China }
\author{Shi-Lin Zhu$^{5,6,7}$}
\email{zhusl@pku.edu.cn} \affiliation{ $^5$School of Physics and
State Key Laboratory of Nuclear Physics and Technology, Peking
University, Beijing 100871, China
\\
$^6$Collaborative Innovation Center of Quantum Matter, Beijing
100871, China
\\
$^7$Center of High Energy Physics, Peking University, Beijing
100871, China }

\date{\today}
\begin{abstract}
In the framework of the color-magnetic interaction, we have
systematically studied the mass splittings of the possible
hidden-charm pentaquarks $qqqc\bar{c}$ ($q=u,d,s$) where the three
light quarks are in a color-octet state. We find that i) the LHCb
$P_c$ states fall in the mass region of the studied system; ii) most
pentaquarks should be broad states since their $S$-wave open-charm
decays are allowed while the lowest state is the $J^P=\frac12^-$
$\Lambda$-like pentaquark with probably the suppressed
$\eta_c\Lambda$ decay mode only; and iii) the $J^P=\frac52^-$ states
do not decay through $S$-wave and their widths are not so broad. The
masses and widths of the two LHCb $P_c$ baryons are compatible with
such pentaquark states. We also explore the hidden-bottom and
$B_c$-like partners of the hidden-charm states and find the possible
existence of the pentaquarks which are lower than the relevant
hadronic molecules.
\end{abstract}


\maketitle
\section{Introduction}\label{sec1}

In 2015, the LHCb Collaboration \cite{Aaij:2015tga} reported two
pentaquark-like resonances $P_c(4380)$ and $P_c(4450)$ in the
process $\Lambda_b^0\to J/\psi K^-p$ with the same decay mode
$J/\psi p$. The decay channel indicates that their minimal quark
content is $nnnc\bar{c}$ ($n=u,d$). The resonance parameters are
$M_{P_c(4380)}=4380\pm8\pm29$ MeV,
$\Gamma_{P_c(4380)}=205\pm18\pm86$ MeV and
$M_{P_c(4450)}=4449.8\pm1.7\pm2.5$ MeV,
$\Gamma_{P_c(4450)}=39\pm5\pm19$ MeV. The preferred angular momenta
are $\frac32$ and $\frac52$, respectively and their $P$ parities are
opposite. Later, the $P_c(4380)$ and $P_c(4450)$ were confirmed by
the reanalysis with a model-independent method \cite{Aaij:2016phn}.
Recently, these two $P_c$ states were also observed in the
$\Lambda_b^0\to J/\psi p\pi^-$ decay \cite{Aaij:2016ymb}.

In fact, the theoretical exploration of the hidden-charm pentaquarks
was performed before the observation of two $P_c$ states by LHCb. In Refs.
\cite{Wu:2010jy,Wu:2010vk}, the authors predicted two
$N^*_{c\bar{c}}$ states and four $\Lambda^*_{c\bar{c}}$ states,
where their masses, decay behaviors and production properties were
given in a coupled-channel unitary approach. Possible molecular states composed of a charmed baryon and an
anticharmed meson were systematically studied with the
one-boson-exchange (OBE) model in Ref. \cite{Yang:2011wz} and the
chiral quark model in Ref. \cite{Wang:2011rga}. More investigations can
be found in Refs.
\cite{Yuan:2012wz,Wu:2012md,Garcia-Recio:2013gaa,Xiao:2013yca,Uchino:2015uha,Huang:2013mua,Wang:2015xwa,Garzon:2015zva}.
In addition, Li and Liu indicated the existence of hidden-charm
pentaquarks by the analysis of a global group structure
\cite{Li:2014gra}.

After the announcement of the $P_c(4380)$ and $P_c(4450)$ states by
LHCb, these two $P_c$ states were interpreted as
$\Sigma_c\bar{D}^*$, $\Sigma_c^*\bar{D}$, or $\Sigma_c^*\bar{D}^*$
molecules
\cite{Chen:2015loa,Roca:2015dva,He:2015cea,Ortega:2016syt,Yamaguchi:2016ote,He:2016pfa,Burns:2015dwa,Huang:2015uda,Chen:2015moa,Chen:2016otp,Shimizu:2016rrd,Shen:2016tzq},
bound states or resonances of charmonium and nucleon
\cite{Meissner:2015mza,Eides:2015dtr,Kahana:2015tkb,Perevalova:2016dln},
diquark-diquark-antiquark states
\cite{Maiani:2015vwa,Anisovich:2015cia,Li:2015gta,Ghosh:2015ksa,Wang:2015ava,Wang:2015epa},
diquark-triquark states \cite{Lebed:2015tna,Zhu:2015bba}, compact
pentaquark states
\cite{Santopinto:2016pkp,Deng:2016rus,Takeuchi:2016ejt}, kinematical
effects due to $\chi_{c1}p$ rescattering \cite{Guo:2015umn}, due to
triangle singularity
\cite{Mikhasenko:2015vca,Liu:2015fea,Guo:2016bkl}, or due to a
$\bar{D}$-soliton \cite{Scoccola:2015nia}, or a bound state of the
colored baryon and meson \cite{Mironov:2015ica}. Their decay and
production properties were studied in Refs.
\cite{Wang:2015qlf,Wang:2015jsa,Wang:2015pcn,Kubarovsky:2015aaa,Cheng:2015cca,Hsiao:2015nna,Karliner:2015voa,Lu:2015fva,Wang:2016vxa,Schmidt:2016cmd,Huang:2016tcr,Kim:2016cxr,Blin:2016dlf,Ali:2016dkf,Wang:2016wxg,Kubarovsky:2016whd,Xiao:2016ogq}.

The observation of these $P_c$ resonances also stimulated the
arguments for more possible pentaquarks
\cite{Lebed:2015dca,Yang:2015bmv,Anisovich:2015zqa,Wang:2016dzu,Chen:2016ryt}.
Productions of another $N^*$ and $\Lambda_{c\bar{c}}^*$ were
discussed in Ref. \cite{Ouyang:2015rre} and Refs.
\cite{Chen:2015sxa,Feijoo:2015kts,Lu:2016roh}, respectively. For the
detailed overview on the hidden-charm pentaquarks, the readers may
refer to Refs. \cite{Chen:2016qju,Chen:2016heh}.

The dynamical calculations of the bound states are relatively easier
if one treats the system as two clusters. The investigation at the
quark level is also simplified when one assumes the existence of
substructures in a five-body system. If a hidden-charm pentaquark
really exists, its spin partners with the same flavor content should
also exist. The existence of substructures certainly results in less
pentaquarks. Needless to say, configurations with various
substructures (baryon-meson, diquark-diquark-antiquark, or
diquark-triquark) lead to different results. From symmetry
consideration, a physical pentaquark state should be a mixture of
all these configurations with various color structures. We here
would like to explore a pentaquark structure without the assumption
of its substructure.

The masses of the $P_c$'s are both above the threshold of $J/\psi
p$. Because the interaction between the $J/\psi$ and nucleon is very
weak, the scattering resonances in this channel are not appropriate
interpretations for the observed $P_c$ states. We focus on the
possible pentaquark configurations where either the three light
quark $qqq$ or the $c\bar{c}$ pair is a color octet state. We
investigate whether the lower $P_c$ state can be assigned as a
tightly bound five-quark state and explore its possible partner
states. Recently, there appeared a preliminary quark model study on
the hidden color-octet $uud$ baryons \cite{Takeuchi:2016ejt}.

In principle, a dynamical calculation for a five-body problem is
needed in order to calculate their masses. In this work, we
calculate their mass splittings with a simple color-magnetic
interaction from the one-gluon-exchange (OGE) potential. For
example, the $\Delta^+$ baryon and the proton have the same quark
content and color structure and their mass difference mainly arises
from the color-magnetic interaction. With the calculated mass
splittings and a reference threshold, one can estimate the
pentaquark masses roughly.

This paper is organized as follows. In Sec. \ref{sec2}, we construct
the $flavor\otimes color\otimes spin$ wave functions of the
hidden-charm pentaquark states and calculate the matrix elements for
the color magnetic interaction in the symmetric limit. Then we
consider the flavor breaking case in Sec. \ref{sec3} and give
numerical results in Sec. \ref{sec4}. In Sec. \ref{sec5}, we explore
the heavier pentaquarks. We discuss our results and summarize in the
final section.

\section{Wave functions and color-magnetic interaction}\label{sec2}

The color-magnetic Hamiltonian reads
\begin{eqnarray}
H&=&\sum_im_i+H_{CM},\nonumber\\
H_{CM}&=&-\sum_{i<j}C_{ij}\lambda_i\cdot\lambda_j\sigma_i\cdot\sigma_j,
\end{eqnarray}
where the $i$-th Gell-Mann matrix $\lambda_i$ should be replaced
with $-\lambda_i^*$ for an antiquark. In the Hamiltonian, $m_i$ is
the effective mass of the $i$-th quark and $C_{ij}\sim\langle
\delta(r_{ij})\rangle/(m_im_j)$ is the effective coupling constant
between the $i$-th quark and the $j$-th quark. The values of the
parameters for light quarks and those for heavy quarks are different
and they will be extracted from the known hadron masses. For the
hidden-charm systems, we have four types of coupling parameters
$C_{qq}$, $C_{qc}$, $C_{q\bar{c}}$, and $C_{c\bar{c}}$ with
$q=u,d,s$. More parameters need to be determined for the other
$qqqQ\bar{Q}$ pentaquarks ($Q=b,c$). In Ref. \cite{Hyodo:2012pm}, we
have estimated the mass of another not-yet-observed but plausible
exotic meson $T_{cc}$ with this simple model. In Refs.
\cite{Wu:2016vtq,Wu:2016gas,Chen:2016ont}, we discussed the mass
splittings for the $QQ\bar{Q}\bar{Q}$, $cs\bar{c}\bar{s}$, and
$QQ\bar{Q}\bar{q}$ systems, respectively.

To calculate the required matrix elements of the color-magnetic
interaction (CMI), we here construct the flavor-color-spin wave
functions of the ground state pentaquark systems. These wave
functions will also be useful in the study of other properties of
the pentaquak states in quark models. In Ref. \cite{Yuan:2012wz}, a
study with the color-magnetic interaction is also involved but the
wave functions are constructed with flavor $SU(4)$ symmetry. Now we
consider flavor $SU(3)$ symmetry and treat the heavy (anti)quark as
a flavor singlet state.

Because the three light quarks $qqq$ must obey Pauli principle, it
is convenient to discuss the constraint with flavor-spin
$SU(6)_{fs}$ symmetry. The three-quark colorless ground baryons
belong to the symmetric $[3]=56$ representation. Its $SU(3)_f\otimes
SU(2)_s$ decomposition gives $(10,4)+(8,2)$ and therefore flavor
singlet baryon in $3\otimes3\otimes3=10+8+8+1$ is forbidden. Now the
color-octet $qqq$ must belong to the mixed $[21]=70$ $SU(6)_{fs}$
representation. The $SU(3)_f\otimes SU(2)_s$ decomposition gives
$(10,2)+(8,4)+(8,2)+(1,2)$. So the flavor singlet pentaquark is
allowed and we have two flavor octets with different spins. There is
no symmetry constraint for the heavy quark pair and one finally gets
three pentaquark decuplets with $J=1/2$, 1/2, and 3/2, three octets
with $J=1/2$, 1/2, and 3/2, four octets with $J=3/2$, 1/2, 3/2, and
5/2, and three singlets with $J=1/2$, 1/2, and 3/2.

\begin{table}[h!]
\begin{tabular}{cccccc}
Multiplet &Space& \multicolumn{2}{c}{Wave function} &
\multicolumn{2}{c}{Wave function}\\\hline\hline
$10_f$ & Color &\multicolumn{2}{c}{$\phi^{MS}$} & \multicolumn{2}{c}{$\phi^{MA}$}\\
& Spin & \multicolumn{2}{c}{$\chi^{MA}$} & \multicolumn{2}{c}{$\chi^{MS}$}\\
$[111]_{cs}$& Flavor & \multicolumn{2}{c}{$F^{S}$} & \multicolumn{2}{c}{$F^{S}$}\\
\hline
$1_f$ & Color &\multicolumn{2}{c}{$\phi^{MS}$} & \multicolumn{2}{c}{$\phi^{MA}$}\\
& Spin & \multicolumn{2}{c}{$\chi^{MS}$} & \multicolumn{2}{c}{$\chi^{MA}$}\\
$[3]_{cs}$& Flavor & \multicolumn{2}{c}{$F^{A}$} & \multicolumn{2}{c}{$F^{A}$}\\
\hline
$8_f(1)$&Color &\multicolumn{2}{c}{$\phi^{MS}$} & \multicolumn{2}{c}{$\phi^{MA}$}\\
& Spin & \multicolumn{2}{c}{$\chi^{S}$} & \multicolumn{2}{c}{$\chi^{S}$}\\
$[21]_{cs}$& Flavor &\multicolumn{2}{c}{$F^{MA}$} & \multicolumn{2}{c}{$F^{MS}$}\\
\hline
$8_f(2)$&Color &\multicolumn{2}{c}{$\phi^{MS}$} & \multicolumn{2}{c}{$\phi^{MA}$}\\
& Spin & $\chi^{MS}$&$\chi^{MA}$ & $\chi^{MS}$&$\chi^{MA}$\\
$[21]_{cs}$& Flavor &$F^{MA}$&$F^{MS}$ & $F^{MS}$&$F^{MA}$\\\hline
\end{tabular}
\caption{Flavor multiplets and wave functions of the colored $qqq$
in different spaces. Young diagrams for the color-spin $SU(6)_{cs}$
are also given in the first column.}\label{combination}
\end{table}

\begin{figure*}[htbp]
\includegraphics[width=400pt]{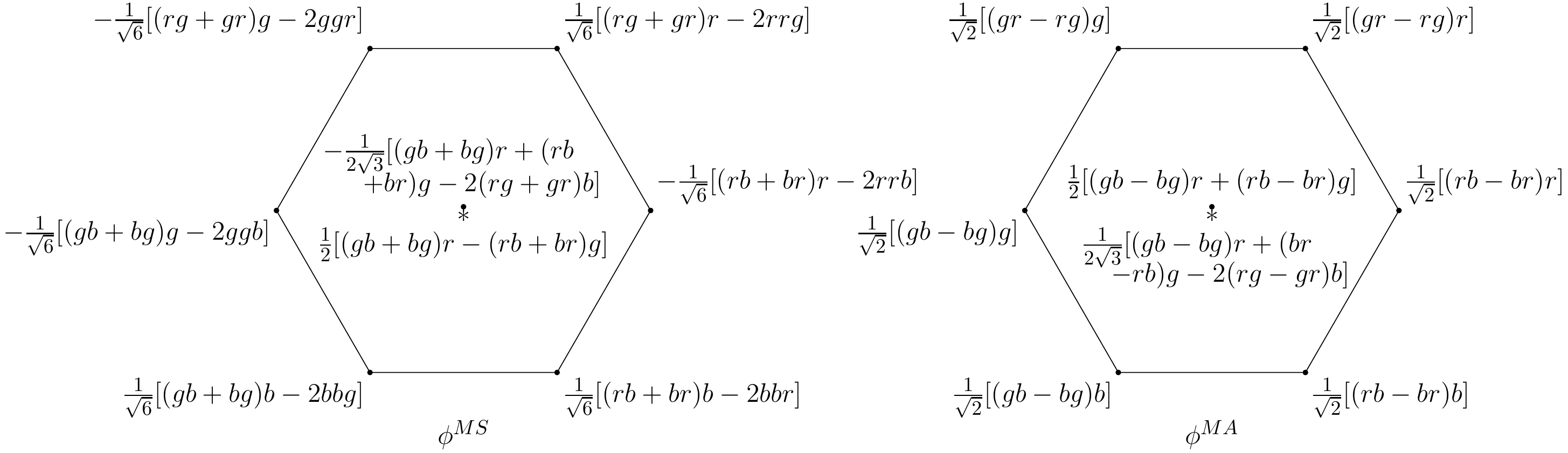}
\caption{Color wave functions of the light
quarks.}\label{qqqcolorwave}
\end{figure*}

\begin{figure}[h!]
\includegraphics[width=150pt]{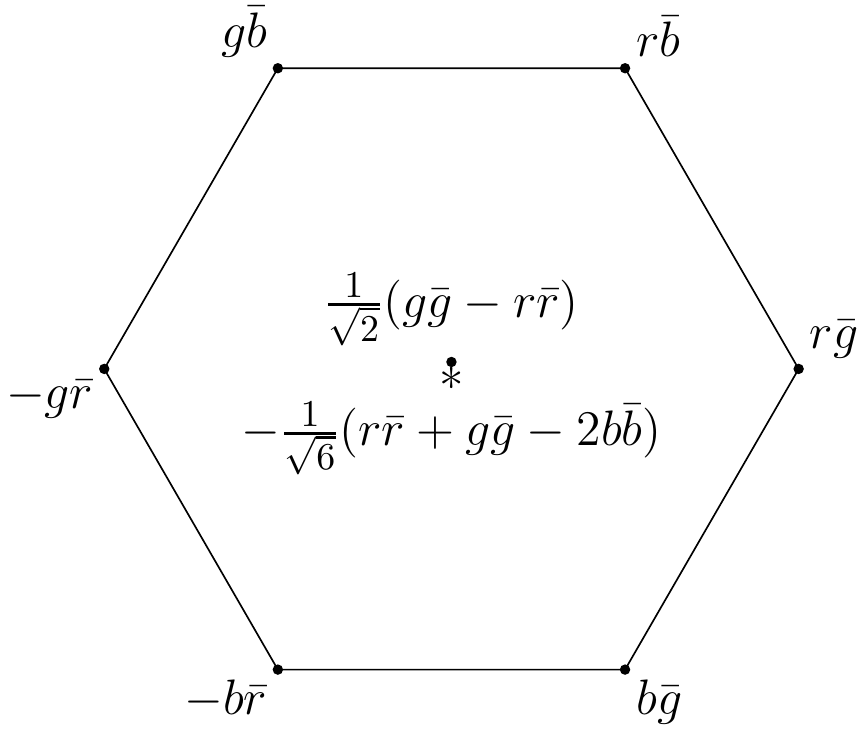}
\caption{Color wave functions of the heavy quark
pair.}\label{ccbarcolorwave}
\end{figure}

To get a totally antisymmetric $qqq$ wave function, we need the
components presented in Tab. \ref{combination} and need to make
appropriate combinations. The notation $MS$ ($MA$) means that the
first two quarks are symmetric (antisymmetric) when they are
exchanged in corresponding space and the superscript $S$ ($A$) means
that the wave function is totally symmetric (antisymmetric). When
one combines the color and spin (or flavor and color, or flavor and
spin), a wave function with a required symmetry is determined by the
relative sign between different components. For example,
$(\phi^{MS}\otimes\chi^{MS}+\phi^{MA}\otimes\chi^{MA})$ is symmetric
for the exchange of color and spin indices simultaneously for any
two quarks. If one uses a minus sign, the wave function is symmetric
only for the first two quarks, i.e. a mixed $MS$ type color-spin
wave function. One has to adopt a self-consistent convention for the
SU(3) Clebsch-Gordan (C.G.) coefficients when constructing the wave
functions. We here take the convention convenient for use
\cite{deSwart:1963pdg,Kaeding:1995vq}. For clarity, we present the
color wave functions of $qqq$ in Fig. \ref{qqqcolorwave} and those
of $c\bar{c}$ in Fig. \ref{ccbarcolorwave}. The flavor octet wave
functions are easy to obtain with the replacements $r\to u$, $g\to
d$, and $b\to s$. The spin wave functions for the spin-half case are
$\chi^{MS}_\uparrow=-\frac{1}{\sqrt6}[(\uparrow\downarrow+\downarrow\uparrow)\uparrow-2\uparrow\uparrow\downarrow]$,
$\chi^{MS}_\downarrow=\frac{1}{\sqrt6}[(\uparrow\downarrow+\downarrow\uparrow)\downarrow-2\downarrow\downarrow\uparrow]$,
$\chi^{MA}_\uparrow=\frac{1}{\sqrt2}(\uparrow\downarrow-\downarrow\uparrow)\uparrow$,
and
$\chi^{MA}_\downarrow=\frac{1}{\sqrt2}(\uparrow\downarrow-\downarrow\uparrow)\downarrow$.
There is no confusion for the totally symmetric flavor and spin wave
functions and we do not show them explicitly. With explicit
calculation, we find the totally antisymmetric wave functions of the
colored $qqq$ states and show them in Tab. \ref{wave-qqq}.

\begin{table}[htbp]
\begin{tabular}{cc}
Multiplet &Flavor-color-spin wave function\\\hline
$10_f$ & $\frac{1}{\sqrt2}[F^{S}\otimes(\phi^{MS}\otimes\chi^{MA}-\phi^{MA}\otimes\chi^{MS})]$\\
$1_f$ & $\frac{1}{\sqrt2}[F^{A}\otimes(\phi^{MS}\otimes\chi^{MS}+\phi^{MA}\otimes\chi^{MA})]$\\
$8_f(1)$&$\frac{1}{\sqrt2}[(F^{MS}\otimes\phi^{MA}-F^{MA}\otimes\phi^{MS})\otimes\chi^{S}]$\\
$8_f(2)$&$\frac12[(F^{MS}\otimes\chi^{MA}+F^{MA}\otimes\chi^{MS})\otimes\phi^{MS}$ \\
&$+(F^{MS}\otimes\chi^{MS}-F^{MA}\otimes\chi^{MA})\otimes\phi^{MA}]$\\\hline
\end{tabular}
\caption{Antisymmetric wave function for a color-octet $qqq$
state.}\label{wave-qqq}
\end{table}

With the above wave functions and the C.G. coefficients of SU(3)
\cite{deSwart:1963pdg,Kaeding:1995vq} and SU(2), one can construct
the pentaquark wave functions. We here only show the color part
\begin{eqnarray}
\phi^{MS,MA}_{penta}&=&\frac{1}{2\sqrt2}\Bigg[-p_C^{MS,MA}(b\bar{r})-n_C^{MS,MA}(b\bar{g})-\Sigma_C^{+MS,MA}(g\bar{r})\nonumber\\&& +\Sigma_C^{-MS,MA}(r\bar{g})-\Xi_C^{0MS,MA}(g\bar{b})+\Xi_C^{-MS,MA}(r\bar{b})\nonumber\\
&&-\frac{1}{\sqrt2}\Sigma_C^{0MS,MA}(g\bar{g}-r\bar{r})+\frac{1}{\sqrt6}\Lambda_C^{MS,MA}(r\bar{r}+g\bar{g}-2b\bar{b})\Bigg],
\end{eqnarray}
where the baryon symbols with the subscript 'C' are borrowed from
flavor octet and represent the color wave functions in Fig.
\ref{qqqcolorwave}. The structure of the full wave functions is the
same as that in Tab. \ref{wave-qqq} by adding a subscript ``penta''
to each wave function.

Since the color-spin interaction is the same for baryons in the same
flavor multiplet in the SU(3) limit, it is enough to consider only
pentaquarks with flavor content $uuuc\bar{c}$ in decuplet,
$uudc\bar{c}$ in octet, and $udsc\bar{c}$ in singlet. After some
calculations we get the results for the $\langle H_{CM}\rangle$ as
follows,
\begin{widetext}
\begin{eqnarray}\label{cmi10}
10_f&:&\langle H_{CM}\rangle=10C_{qq}+2C_{c\bar{c}}, \text{  for  } \Big(S_{c\bar{c}}=0,J=\frac12\Big)\nonumber\\
&&\langle H_{CM}\rangle=10C_{qq}-\frac{2}{3}C_{c\bar{c}}-\frac{20}{3}(C_{qc}-C_{q\bar{c}}),  \text{  for  } \Big(S_{c\bar{c}}=1,J=\frac12\Big)\nonumber\\
&&\langle
H_{CM}\rangle=10C_{qq}-\frac{2}{3}C_{c\bar{c}}+\frac{10}{3}(C_{qc}-C_{q\bar{c}}),
\text{  for  } \Big(S_{c\bar{c}}=1,J=\frac32\Big),
\end{eqnarray}
\begin{eqnarray}\label{cmi1}
1_f&:&\langle H_{CM}\rangle=-14C_{qq}+2C_{c\bar{c}}, \text{  for  }\Big(S_{c\bar{c}}=0,J=\frac12\Big)\nonumber\\
&&\langle H_{CM}\rangle=-14C_{qq}-\frac{2}{3}C_{c\bar{c}}-\frac{4}{3}(C_{qc}+11C_{q\bar{c}}),  \text{  for  }\Big(S_{c\bar{c}}=1,J=\frac12\Big)\nonumber\\
&&\langle
H_{CM}\rangle=-14C_{qq}-\frac{2}{3}C_{c\bar{c}}+\frac{2}{3}(C_{qc}+11C_{q\bar{c}}),
\text{  for  } \Big(S_{c\bar{c}}=1,J=\frac32\Big),
\end{eqnarray}
\begin{eqnarray}\label{cmi81}
8_f(1)&:&\langle H_{CM}\rangle=2C_{qq}+2C_{c\bar{c}}, \text{  for  } \Big(S_{c\bar{c}}=0,J=\frac32\Big)\nonumber\\
&&\langle H_{CM}\rangle=2C_{qq}-\frac{2}{3}C_{c\bar{c}}-10(C_{qc}+C_{q\bar{c}}),  \text{  for  } \Big(S_{c\bar{c}}=1,J=\frac12\Big)\nonumber\\
&&\langle H_{CM}\rangle=2C_{qq}-\frac{2}{3}C_{c\bar{c}}-4(C_{qc}+C_{q\bar{c}}),  \text{  for  } \Big(S_{c\bar{c}}=1,J=\frac32\Big)\nonumber\\
&&\langle
H_{CM}\rangle=2C_{qq}-\frac{2}{3}C_{c\bar{c}}+6(C_{qc}+C_{q\bar{c}}),
\text{  for  } \Big(S_{c\bar{c}}=1,J=\frac52\Big),
\end{eqnarray}
\begin{eqnarray}\label{cmi82}
8_f(2)&:&\langle H_{CM}\rangle=-2C_{qq}+2C_{c\bar{c}}, \text{  for  } \Big(S_{c\bar{c}}=0,J=\frac12\Big)\nonumber\\
&&\langle H_{CM}\rangle=-2C_{qq}-\frac{2}{3}C_{c\bar{c}}-4(C_{qc}+C_{q\bar{c}}),  \text{  for  } \Big(S_{c\bar{c}}=1,J=\frac12\Big)\nonumber\\
&&\langle
H_{CM}\rangle=-2C_{qq}-\frac{2}{3}C_{c\bar{c}}+2(C_{qc}+C_{q\bar{c}}),
\text{  for  } \Big(S_{c\bar{c}}=1,J=\frac32\Big).
\end{eqnarray}
\end{widetext}
One may confirm the part for $C_{qq}$ with the formula
\cite{Oka:2012zz,Maeda:2015hxa}
\begin{eqnarray}
\Big\langle\sum_{i<j}(\lambda_i\cdot\lambda_j)(\sigma_i\cdot\sigma_j)\Big\rangle
=- \Big[8N+\frac43S(S+1)+2C_2[SU(3)_c]-4C_2[SU(6)_{cs}]\Big],
\end{eqnarray}
where $N=3,S=\frac12$ or $\frac32$, and $C_2[SU(g)]$ is the
quadratic Casimir operator specified by the Young diagram
$[f_1,...,f_g]$
\begin{eqnarray}
C_2[SU(g)]=\frac12\left[\sum_i f_i(f_i-2i+g+1)-\frac{N^2}{g}\right].
\end{eqnarray}
The Young diagram for color symmetry is $[21]_c$ and those for
color-spin $SU(6)_{cs}$ symmetry can be found in Tab.
\ref{combination}. The part for $C_{c\bar{c}}$ can be verified with
the formula
\begin{eqnarray}
\Big\langle(\lambda_4\cdot\lambda_5)(\sigma_4\cdot\sigma_5)\Big\rangle
=4\Big[C_2[SU(3)_c]-\frac83\Big]\Big[S_{c\bar{c}}(S_{c\bar{c}}+1)-\frac32\Big].
\end{eqnarray}
However, it is problematic to discuss mass splittings for
pentaquarks with Eqs. (\ref{cmi10})--(\ref{cmi82}) directly because
of violations of the heavy quark spin symmetry (HQSS) and the flavor
SU(3) symmetry.

The heavy quark symmetry is strict in the limit $m_c\to \infty$,
which leads to the irrelevance of the heavy quark spin (and flavor)
for the interaction between a heavy quark and a light quark. In this
limit, the interaction within  the heavy quark pair is also
irrelevant with their spin (but not the flavor) and the spin-flip
between the $S_{c\bar{c}}=0$ case and the $S_{c\bar{c}}=1$ case is
suppressed. The color-magnetic interaction obviously violates HQSS.
This means that only $C_{qq}$ terms in Eqs.
(\ref{cmi10})--(\ref{cmi82}) are important in the heavy quark limit
and there are four degenerate multiplets with the mass ordering:
$10_f$, $8_f(1)$, $8_f(2)$, and $1_f$ from high to low.

After the heavy quark mass correction is included, all the terms
involving $m_c$ in Eqs. (\ref{cmi10})--(\ref{cmi82})  contribute.
Since now the spin-flip between the $S_{c\bar{c}}=0$ case and the
$S_{c\bar{c}}=1$ case is considered, the mixing between states with
the same $J$ occurs, which results from the term proportional to
$1/{m_cm_q}$. Then one should determine the final $\langle
H_{CM}\rangle$'s for mixed states by diagonalizing the specified
matrix.

Usually, the flavor mixing between different multiplet
representations occurs once the symmetry breaking is considered. In
the present case, even in the $SU(3)_f$ limit, the mixing between
the two octets is nonvanishing, which complicates the color magnetic
interactions. To be convenient, we now collect the averages of the
CMI in a matrix form for the $nnnc\bar{c}$ ($n=u,d$) and
$sssc\bar{c}$ cases. For the other cases containing the $s$ quark,
we show results in the next section.

For $I=\frac32$ $nnnc\bar{c}$ states (3 baryons in $10_f$),
\begin{eqnarray}
\langle H_{CM}\rangle_{J=\frac32}&=&10C_{nn}+\frac{10}{3}(C_{nc}-C_{n\bar{c}})-\frac23C_{c\bar{c}},\nonumber\\
\langle H_{CM}\rangle_{J=\frac12}&=&\left(\begin{array}{cc}
10C_{nn}-\frac{20}{3}(C_{nc}-C_{n\bar{c}})-\frac23C_{c\bar{c}}&\frac{10}{\sqrt3}(C_{nc}+C_{n\bar{c}})\\
&10C_{nn}+2C_{c\bar{c}}\end{array}\right).
\end{eqnarray}
One gets similar expressions for the $I=0$ $sssc\bar{c}$ states (3
baryons in $10_f$) by replacing $n$ with $s$.

For $I=1/2$ $nnnc\bar{c}$ states (7 baryons in $8_f$), the results
read
\begin{eqnarray}
\langle H_{CM}\rangle_{J=\frac52}&=&2C_{nn}+6(C_{nc}+C_{n\bar{c}})-\frac{2}{3}C_{c\bar{c}},\nonumber\\
\langle H_{CM}\rangle_{J=\frac32}&=&\left(\begin{array}{cccc}
2C_{nn}-4(C_{nc}+C_{n{\bar{c}}})-\frac23C_{c\bar{c}}&2\sqrt{15}(C_{nc}-C_{n{\bar{c}}})&-\frac{2\sqrt{10}}{3}(C_{nc}-4C_{n\bar{c}})\\
&2(C_{nn}+C_{c\bar{c}})&\frac{2\sqrt{6}}{3}(C_{nc}+4C_{n\bar{c}})\\
&&-2C_{nn}+2(C_{nc}+C_{n{\bar{c}}})-\frac23C_{c\bar{c}}
\end{array}\right),\nonumber\\
\langle H_{CM}\rangle_{J=\frac12}&=&\left(\begin{array}{cccc}
2C_{nn}-10(C_{nc}+C_{n{\bar{c}}})-\frac23C_{c\bar{c}}&-\frac{4}{\sqrt3}(C_{nc}+4C_{n\bar{c}})&-\frac43(C_{nc}-4C_{n\bar{c}})\\
&2(-C_{nn}+C_{c\bar{c}})&2\sqrt3(C_{nc}-C_{n\bar{c}})\\
&&-2C_{nn}-4(C_{nc}+C_{n{\bar{c}}})-\frac23C_{c\bar{c}}
\end{array}\right),
\end{eqnarray}
where the bases for $J=\frac32$ and $J=\frac12$ are
$\Big(8_f(1)^{[S_{c\bar{c}}=1]}$, $8_f(1)^{[S_{c\bar{c}}=0]}$,
$8_f(2)^{[S_{c\bar{c}}=1]}\Big)^T$ and
$\Big(8_f(1)^{[S_{c\bar{c}}=1]}$, $8_f(2)^{[S_{c\bar{c}}=0]}$,
$8_f(2)^{[S_{c\bar{c}}=1]}\Big)^T$, respectively.

\section{$SU(3)_f$ breaking}\label{sec3}

Up to now, we have not considered the $SU(3)_f$ breaking. Once the
mass difference between the strange quark and the $u,d$ quarks is
included, the general mixing between flavor multiplets appears. Such
an effect is included in the color-magnetic interaction and we now
discuss this case. The systems we need to consider additionally are
$nnsc\bar{c}$ and $ssnc\bar{c}$. They are classified into two
categories according to the symmetry for the first two quarks in
flavor space: symmetric $nnsc\bar{c}$ ($I=1$) and $ssnc\bar{c}$
($I=0$) and antisymmetric $nnsc\bar{c}$ ($I=0$). In the following,
we use the symbol like $[(qqq')^{MA}_{MS}(c\bar{c})^0_{8}]^J$ to
denote the base states. In this example, the subscript $MS$ means
that the color representation for the $(qqq')$ is $8^{MS}$ and the
color wave function is $\phi^{MS}$. The subscript 8 is the color
representation for the $(c\bar{c})$. The superscript $MA$ means that
the spin wave function for the $(qqq')$ is $\chi^{MA}$ and the spin
is 1/2. The superscript $0$ ($J$) indicates the spin of the
$(c\bar{c})$ (pentaquark).

The calculation method can be found in Refs.
\cite{Hogaasen:2004pm,Buccella:2006fn}. We first give the results
for the symmetric category. For the case $J=\frac52$,
\begin{eqnarray}\label{qq1qpHcmJ5}
\langle
H_{CM}\rangle_{J=\frac52}=\frac23(4C_{12}-C_{13}+7C_{14}+2C_{15}+2C_{34}+7C_{35}-C_{45}).
\end{eqnarray}
There is only one base state
$[(qqq')^S_{MA}(c\bar{c})^1_8]^{\frac52}$. For the case $J=\frac32$,
we use the base vector
$\Big([(qqq')^S_{MA}(c\bar{c})^1_8]^{\frac32}$,
$[(qqq')^S_{MA}(c\bar{c})^0_8]^{\frac32}$,
$[(qqq')^{MS}_{MA}(c\bar{c})^1_8]^{\frac32}$,
$[(qqq')^{MA}_{MS}(c\bar{c})^0_8]^{\frac32}\Big)^T$ and get
\begin{eqnarray}\label{qq1qpHcmJ3}
\langle H_{CM}\rangle_{J=\frac32}=\left(\begin{array}{cccc}
\frac29(3\mu-2\alpha-2\gamma)&\frac29\sqrt{15}(\beta+\delta)&-\frac{2\sqrt{5}}{9}(\alpha-2\gamma)&\frac{\sqrt{5}}{21}(13\alpha-15\beta)\\
&\frac23(8\lambda-4\nu-7\mu)&\frac{2\sqrt{3}}{9}(\beta-2\delta)&\frac{\sqrt{3}}{21}(15\alpha-13\beta)\\
&&\frac29(3\nu+2\alpha-\gamma)&\frac{1}{21}(42\mu-42\nu+13\alpha-15\beta)\\
&&&\frac{1}{21}(14\lambda+13\gamma+15\delta)
\end{array}\right),
\end{eqnarray}
where $\alpha=7C_{14}+2C_{15}$, $\beta=7C_{14}-2C_{15}$,
$\gamma=2C_{34}+7C_{35}$, $\delta=2C_{34}-7C_{35}$,
$\mu=4C_{12}-C_{13}-C_{45}$, $\nu=4C_{12}+2C_{13}-C_{45}$, and
$\lambda=6C_{12}-C_{45}$. For the case $J=\frac12$, we have
\begin{eqnarray}\label{qq1qpHcmJ1}
\langle H_{CM}\rangle_{J=\frac12} =\left(\begin{array}{ccccc}
\frac29(3\mu-5\alpha-5\gamma)&\frac{2\sqrt{2}}{9}(-\alpha+2\gamma)&\frac{2\sqrt{6}}{9}(-\beta+2\delta)&\frac{\sqrt{2}}{21}(13\alpha-15\beta)
&-\frac{\sqrt{6}}{21}(15\alpha-13\beta)\\
&\frac29(3\nu-4\alpha+2\gamma)&\frac{2\sqrt{3}}{9}(2\beta-\delta)&\frac{2}{21}\left(\begin{array}{c}21\mu-21\nu\\-13\alpha+15\beta\end{array}\right)&-\frac{\sqrt{3}}{21}(15\alpha-13\beta)\\
&&\frac23(8\lambda-8\mu-3\nu)&-\frac{\sqrt{3}}{21}(15\alpha-13\beta)&2(\mu-\nu)\\
&&&\frac{2}{21}(7\lambda-13\gamma-15\delta)&\frac{\sqrt{3}}{21}(15\gamma+13\delta)\\
&&&&2(2C_{12}+C_{45})
\end{array}\right)\nonumber\\
\end{eqnarray}
with the base vector $\Big([(qqq')^S_{MA}(c\bar{c})^1_8]^{\frac12}$,
$[(qqq')^{MS}_{MA}(c\bar{c})^1_8]^{\frac12}$,
$[(qqq')^{MS}_{MA}(c\bar{c})^0_8]^{\frac12}$,
$[(qqq')^{MA}_{MS}(c\bar{c})^1_8]^{\frac12}$,
$[(qqq')^{MA}_{MS}(c\bar{c})^0_8]^{\frac32}\Big)^T$.

Now we present the results for the antisymmetric category. For the
case $J=\frac52$, the base state is
$[(qqq')^S_{MS}(c\bar{c})^1_8]^{\frac52}$ and the matrix element is
\begin{eqnarray}\label{qq0qpHcmJ5}
\langle
H_{CM}\rangle_{J=\frac52}=\frac23(-2C_{12}+5C_{13}+5C_{14}+10C_{15}+4C_{34}-C_{35}-C_{45}).
\end{eqnarray}
For the $J=\frac32$ case, we use the base vector
$\Big([(qqq')^S_{MS}(c\bar{c})^1_8]^{\frac32}$,
$[(qqq')^S_{MS}(c\bar{c})^0_8]^{\frac32}$,
$[(qqq')^{MS}_{MS}(c\bar{c})^1_8]^{\frac32}$,
$[(qqq')^{MA}_{MA}(c\bar{c})^0_8]^{\frac32}\Big)^T$ and the obtained
matrix is
\begin{eqnarray}\label{qq0qpHcmJ3}
\langle H_{CM}\rangle_{J=\frac32} =\left(\begin{array}{cccc}
-\frac29(3\mu'+10\alpha'+2\delta')&\frac{2\sqrt{15}}{9}(5\beta'+\gamma')&\frac{2\sqrt{5}}{9}(-5\alpha'+2\delta')&\frac{\sqrt{5}}{3}(\alpha'-3\beta')\\
&\frac{2}{15}(11\mu'+8\nu'-4\lambda')&\frac{2\sqrt{3}}{9}(5\beta'-2\gamma')&\frac{\sqrt{3}}{3}(3\alpha'-\beta')\\
&&\frac29(10\alpha'-\delta'-3\nu')&\frac{1}{15}(6\mu'-6\nu'+5\alpha'-15\beta')\\
&&&-\frac16(4\lambda'-15\gamma'+13\delta')
\end{array}\right),
\end{eqnarray}
where $\alpha'=C_{14}+2C_{15}$, $\beta'=C_{14}-2C_{15}$,
$\gamma'=4C_{34}+C_{35}$, $\delta'=4C_{34}-C_{35}$,
$\mu'=2C_{12}-5C_{13}+C_{45}$, $\nu'=2C_{12}+10C_{13}+C_{45}$, and
$\lambda'=12C_{12}+C_{45}$. For the $J=\frac12$ case, we have
\begin{eqnarray}\label{qq0qpHcmJ1}
&&\langle H_{CM}\rangle_{J=\frac12}=-\frac29\times\nonumber\\
&&\left(\begin{array}{ccccc}
(3\mu'+25\alpha'+5\delta')&\sqrt{2}(5\alpha'-2\delta')&\sqrt{6}(5\beta'-2\gamma')&\frac{3}{\sqrt{2}}(3\beta'-\alpha')&\frac{3\sqrt{6}}{2}(3\alpha'-\beta')\\
&(20\alpha'-2\delta'+3\nu')&-\sqrt{3}(10\beta'-\gamma')&\frac35\left(\begin{array}{c}3\nu'-3\mu'\\+5\alpha'-15\beta'\end{array}\right)&\frac{3\sqrt{3}}{2}(3\alpha'-\beta')\\
&&\frac35(4\lambda'-16\mu'-3\nu')&\frac{3\sqrt{3}}{2}(3\alpha'-\beta')&\frac95(\nu'-\mu')\\
&&&\frac32(2\lambda'+15\gamma'-13\delta')&\frac{3\sqrt{3}}{4}(13\gamma'-15\delta')\\
&&&&\frac35(9\lambda'-16\mu'-8\nu')
\end{array}\right)\nonumber\\
\end{eqnarray}
with the base vector $\Big([(qqq')^S_{MS}(c\bar{c})^1_8]^{\frac12}$,
$[(qqq')^{MS}_{MS}(c\bar{c})^1_8]^{\frac12}$,
$[(qqq')^{MS}_{MS}(c\bar{c})^0_8]^{\frac12}$,
$[(qqq')^{MA}_{MA}(c\bar{c})^1_8]^{\frac12}$,
$[(qqq')^{MA}_{MA}(c\bar{c})^0_8]^{\frac32}\Big)^T$.

One may use the matrices in this section to numerically reproduce
the $\langle H_{CM}\rangle$'s for the $nnnc\bar{c}$ systems after
diagonalization. For the $J=\frac52$ case, both Eq.
(\ref{qq1qpHcmJ5}) and Eq. (\ref{qq0qpHcmJ5}) give the same formula
and thus the same result when $q'=q=n$. For the case $J=\frac32$,
Eq. (\ref{qq1qpHcmJ3}) and Eq. (\ref{qq0qpHcmJ3}) result in
different eigenvalues by assuming $q'=q=n$. However, one finds that
the common numbers of the two sets of eigenvalues are just the
results for the $I=\frac12$ $nnnc\bar{c}$ systems. The remaining
value given by Eq. (\ref{qq1qpHcmJ3}) is the result for the
$I=\frac32$ $nnnc\bar{c}$ system while that given by Eq.
(\ref{qq0qpHcmJ3}) can be thought as a forbidden number because of
the Pauli principle. The $J=\frac12$ case has similar features with
the $J=\frac32$ case. Probably these features can be used to
simplify the calculation for multiquark systems.

\section{Numerical results for the hidden-charm systems}\label{sec4}

By calculating the CMI matrix elements for ground state baryons and
mesons \cite{Olive:2016xmw}, we extract the effective coupling
parameters presented in Tab. \ref{parameter}. In determining
$C_{cn}$, one may also use the mass difference between $\Lambda_c$
and $\Sigma_c$. The resulting pentaquark masses would be around 10
MeV lower, which is a not a large number in the present method of
estimation. Since we also discuss hidden-bottom and $B_c$-like
pentaquark states, Tab. \ref{parameter} displays relevant
parameters, too. Because there is no experimental data for the
$B_c^*$ meson, we determine the value of $C_{b\bar{c}}$ to be 3.3
MeV from a quark model calculation $m_{B_c^*}- m_{B_c}= 70$ MeV
\cite{Godfrey:1985xj}.

\begin{table}[!h]
\caption{The effective coupling parameters extracted from the mass
differences between ground hadrons.}\label{parameter} \centering
\begin{tabular}{cccccc}
\hline \hline
Hadron&CMI&Hadron&CMI&Parameter(MeV)\\
\hline
$N$&$-8C_{nn}$&$\Delta$&$8C_{nn}$&$C_{nn}=18.4$\\
$\Sigma$&$\frac{8}{3}C_{nn}-\frac{32}{3}C_{n s}$&$\Sigma^*$&$\frac{8}{3}C_{nn}+\frac{16}{3}C_{n s}$&$C_{n s}=12.4$\\
$\Xi^0$&$\frac{8}{3}(C_{ss}-4C_{n s})$&$\Xi^{*0}$&$\frac{8}{3}(C_{ss}+C_{n s})$&\\
$\Omega$&8$C_{ss}$&&&$C_{ss}=6.5$\\
$\Lambda$&$-8C_{nn}$\\
$D$&$-16C_{c\bar{n}}$&$D^{*}$&$\frac{16}{3}C_{c\bar{n}}$&$C_{c\bar{n}}=6.7$\\
$D_s$&$-16C_{c\bar{s}}$&$D_{s}^{*}$&$\frac{16}{3}C_{c\bar{s}}$&$C_{c\bar{s}}$=6.7\\
$B$&$-16C_{b\bar{n}}$&$B^{*}$&$\frac{16}{3}C_{b\bar{n}}$&$C_{b\bar{n}}$=2.1\\
$B_s$&$-16C_{b\bar{s}}$&$B^{*}$&$\frac{16}{3}C_{b\bar{s}}$&$C_{b\bar{s}}$=2.3\\
$\eta_{c}$&$-16C_{c\bar{c}}$&$J/\psi$&$\frac{16}{3}C_{c\bar{c}}$&$C_{c\bar{c}}=5.3$\\
$\eta_{b}$&$-16C_{b\bar{b}}$&$\Upsilon$&$\frac{16}{3}C_{b\bar{b}}$&$C_{b\bar{b}}=2.9$\\
$\Sigma_{c}$&$\frac{8}{3}C_{nn}-\frac{32}{3}C_{cn}$&$\Sigma_{c}^*$&$\frac{8}{3}C_{nn}+\frac{16}{3}C_{cn}$&$C_{cn}=4.0$\\
$\Xi'_{c}$&$\frac{8}{3}C_{n s}-\frac{16}{3}C_{cn}-\frac{16}{3}C_{cs}$&$\Xi_{c}^*$&$\frac{8}{3}C_{n s}+\frac{8}{3}C_{cn}+\frac{8}{3}C_{cs}$&$C_{cs}=4.8$\\
$\Sigma_{b}$&$\frac{8}{3}C_{nn}-\frac{32}{3}C_{bn}$&$\Sigma_{b}^{*}$&$\frac{8}{3}C_{nn}+\frac{16}{3}C_{bn}$&$C_{bn}=1.3$\\
$\Xi'_{b}$&$\frac{8}{3}C_{n
s}-\frac{16}{3}C_{bn}-\frac{16}{3}C_{bs}$&$\Xi_b^*$&
$\frac{8}{3}C_{n s}+\frac{8}{3}C_{bn}+\frac{8}{3}C_{bs}$&$C_{bs}=1.2$\\
\hline
\end{tabular}
\end{table}

In the simple model used in the present study, the mass splittings
of the pentaquark states mainly rely on the coupling parameters. The
masses can be roughly estimated with the formula $M=\sum_i
m_i+\langle H_{CM}\rangle$ or from a reference mass
$M=M_{ref}-\langle H_{CM}\rangle_{ref}+\langle H_{CM}\rangle$. In
the latter scheme, the reference system has the same quark content
with the pentaquark system and the dependence on the effective quark
masses is partially canceled. We will show results in both schemes.
Here, the effective quark masses are $m_n= 361.8$ MeV, $m_s=540.4$
MeV, $m_c=1724.8$ MeV, and $m_b=5052.9$ MeV, which are also
extracted from the ground hadrons. In Ref. \cite{Wu:2016gas}, it is
illustrated that these effective quark masses result in
overestimated hadron masses and one may treat the values as
theoretical upper limits. Then we mainly focus on the second scheme
and use various meson-baryon thresholds as reference masses.

\subsection{The $nnnc\bar{c}$ system}

In this case, there are two types of thresholds we may use:
(charmonium)+(light baryon) and (charmed baryon)+(charmed meson).
However, the CMI matrix elements $\langle H_{CM}\rangle_{(J/\psi
p)}=-119$ MeV and $\langle H_{CM}\rangle_{(\Sigma_c\bar{D})}=-102$
MeV are not consistent with the thresholds (4035 MeV for $J/\psi p$
and 4320 MeV for $\Sigma_c\bar{D}$). This indicates that one cannot
eliminate completely the quark mass effects with the reference mass
scheme. The reason is that the model does not involve dynamics and
the contributions from the other terms in the potential model are
related with the system structure. For example, the additional
kinetic energy can probably shift the estimated mass to a more
physical value \cite{Park:2015nha}. To understand which threshold is
more reasonable, one needs detailed calculation in a future work.
Here, we estimate pentaquark masses with both types of thresholds.
The numerical results are given in Tab. \ref{mass-nnnccbar}. The use
of the $J/\psi N$, $J/\psi\Delta$, $\eta_c N$, and $\eta_c\Delta$
thresholds gives similar values and that of $\Sigma_c\bar{D}$,
$\Sigma_c\bar{D}^*$, $\Sigma_c^*\bar{D}$, and $\Sigma_c^*\bar{D}^*$
gives similar values. The reference threshold $\Lambda_c\bar{D}$ or
$\Lambda_c\bar{D}^*$ results in around 10 MeV lower masses than
$\Sigma_c\bar{D}$ does. We do not present these results in the
table.

\begin{table}[!h]
\caption{Calculated CMI's and estimated pentaquark masses of the
$nnnc\bar{c}$ systems in units of MeV. The masses in the forth
column are calculated with the effective quark masses and are
theoretical upper limits.}\label{mass-nnnccbar}
\begin{tabular}{c|ccccc}\hline
\multicolumn{6}{c}{$nnnc\bar{c}$ ($I=\frac32$)} \\\hline\hline
$J^{P}$ & $\langle H_{CM} \rangle$ &Eigenvalue
&Mass&$(J/\psi\Delta)$ &$(\Sigma_c\bar{D})$\\\hline
$\frac32^{-}$ &171.5&171.5&4706.5&4325.5&4591.1\\
$\frac12^{-}$ &$\left(\begin{array}{cc}198.5&61.8\\61.8&194.6\end{array}\right)$&$\left(\begin{array}{c}258.3\\134.7\end{array}\right)$&$\left(\begin{array}{c}4793.3\\4669.7\end{array}\right)$&$\left(\begin{array}{c}4412.4\\4288.8\end{array}\right)$&$\left(\begin{array}{c}4677.9\\4554.3\end{array}\right)$\\
\hline\hline\multicolumn{6}{c}{$nnnc\bar{c}$ ($I=\frac12$)}
\\\hline\hline $J^{P}$ & $\langle H_{CM} \rangle$ &Eigenvalue
&Mass&$(J/\psi N)$&$(\Sigma_c\bar{D})$\\\hline
$\frac52^{-}$ &97.5&97.5&4632.5&4251.6&4517.1\\
$\frac32^{-}$ &$\left(\begin{array}{ccc}-9.5&-20.9&48.1\\-20.9&47.4&50.3\\48.1&50.3&-18.9\end{array}\right)$&$\left(\begin{array}{c}-83.1\\74.8\\27.3\end{array}\right)$&$\left(\begin{array}{c}4451.9\\4609.8\\4562.3\end{array}\right)$&$\left(\begin{array}{c}4071.0\\4228.9\\4181.4\end{array}\right)$&$\left(\begin{array}{c}4336.5\\4494.4\\4446.9\end{array}\right)$\\
$\frac12^{-}$ &$\left(\begin{array}{ccc}-73.7&-71.1&30.4\\-71.1&-26.2&-9.4\\30.4&-9.4&-83.1\end{array}\right)$&$\left(\begin{array}{c}-133.0\\-80.8\\30.8\end{array}\right)$&$\left(\begin{array}{c}4402.0\\4454.2\\4565.8\end{array}\right)$&$\left(\begin{array}{c}4021.1\\4073.3\\4184.9\end{array}\right)$&$\left(\begin{array}{c}4286.6\\4338.8\\4450.4\end{array}\right)$\\
\hline
\end{tabular}
\end{table}

\begin{figure}[!h]
\begin{tabular}{ccc}
\includegraphics[width=220pt]{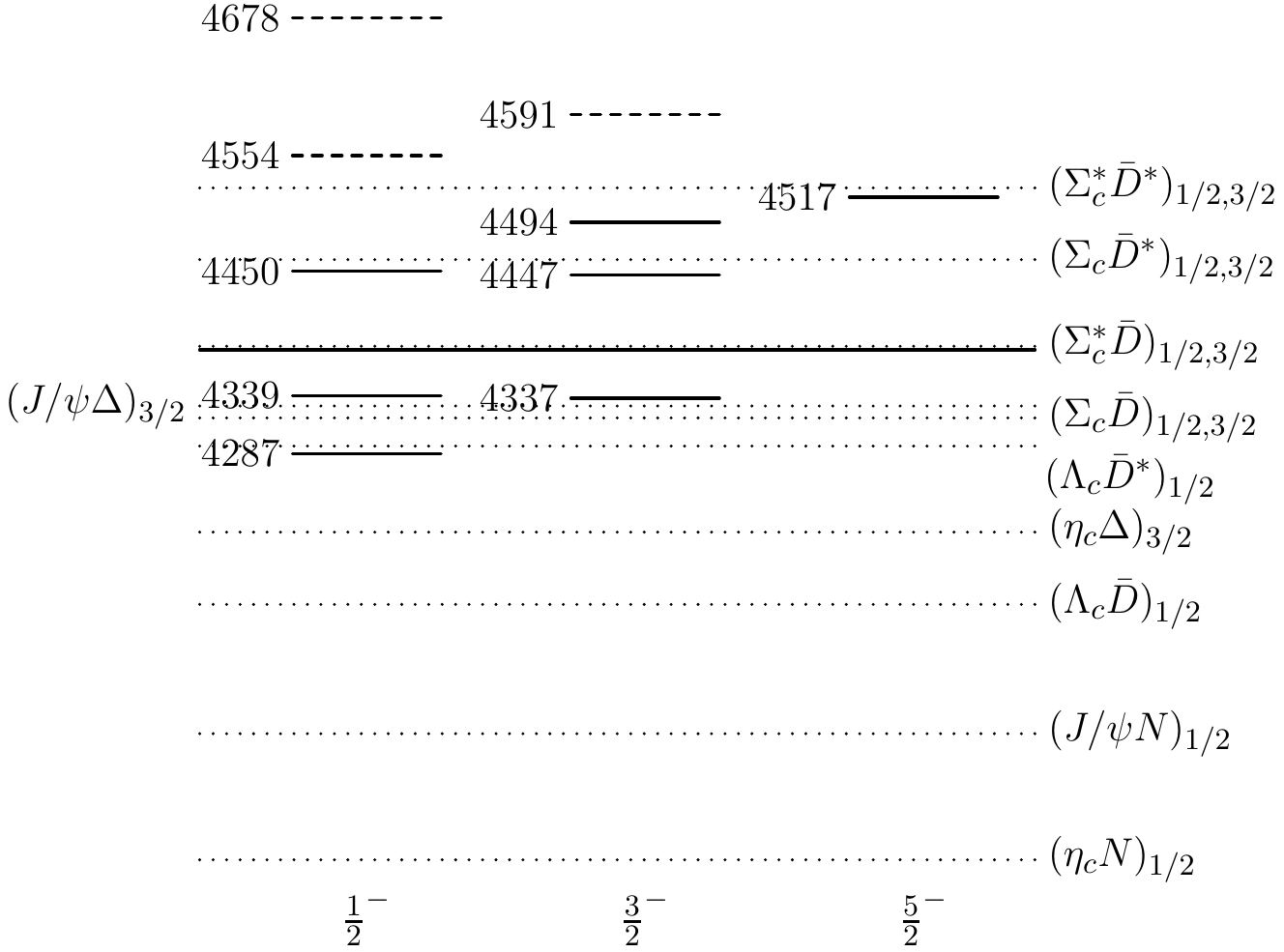}&$\qquad$&
\includegraphics[width=220pt]{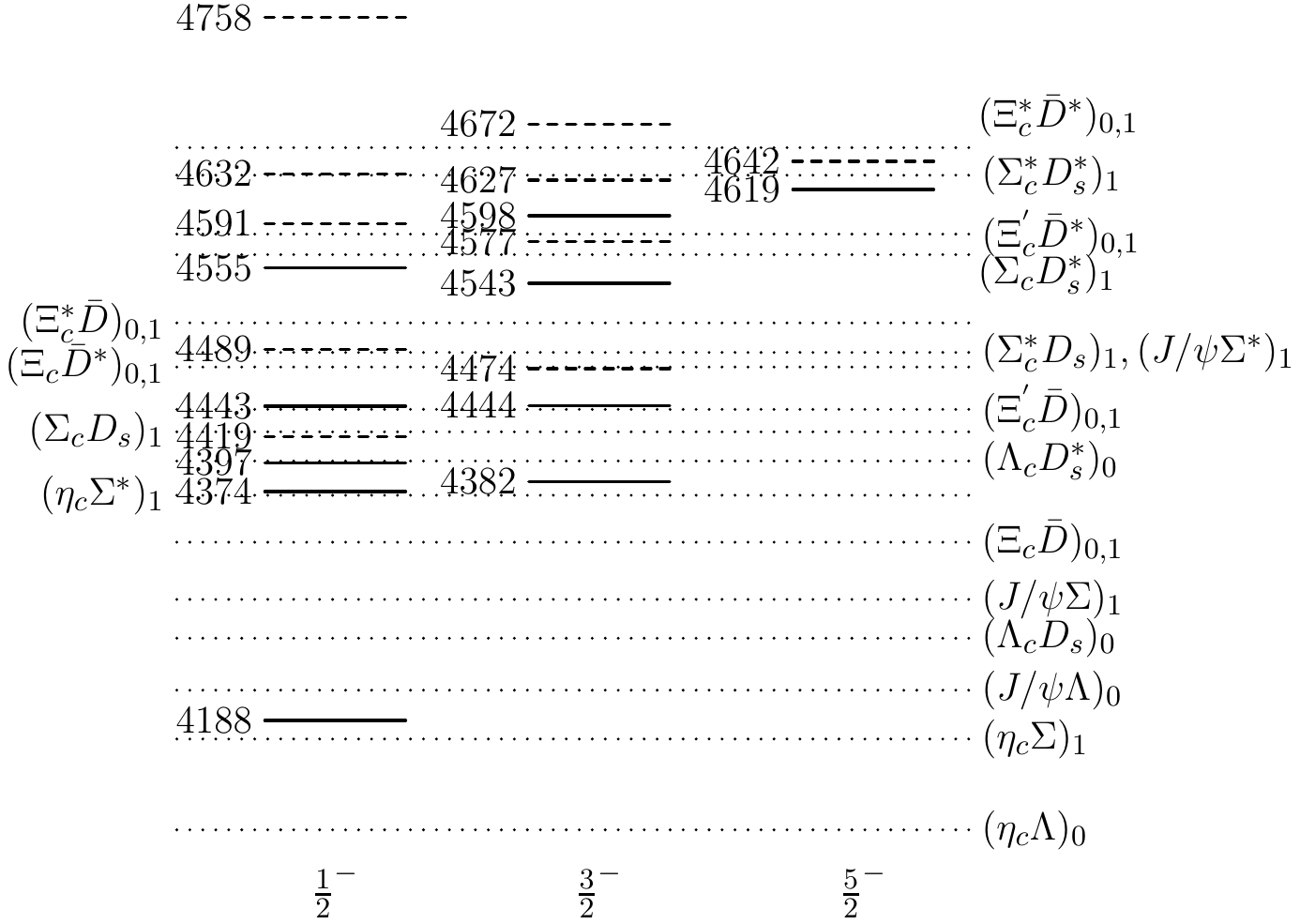}\\
(a) $I=\frac32$ (dashed) and $I=\frac12$ (solid) $nnnc\bar{c}$ states &&(b) $I=1$ (dashed) and $I=0$ (solid) $nnsc\bar{c}$ states\\
&&\\
\includegraphics[width=220pt]{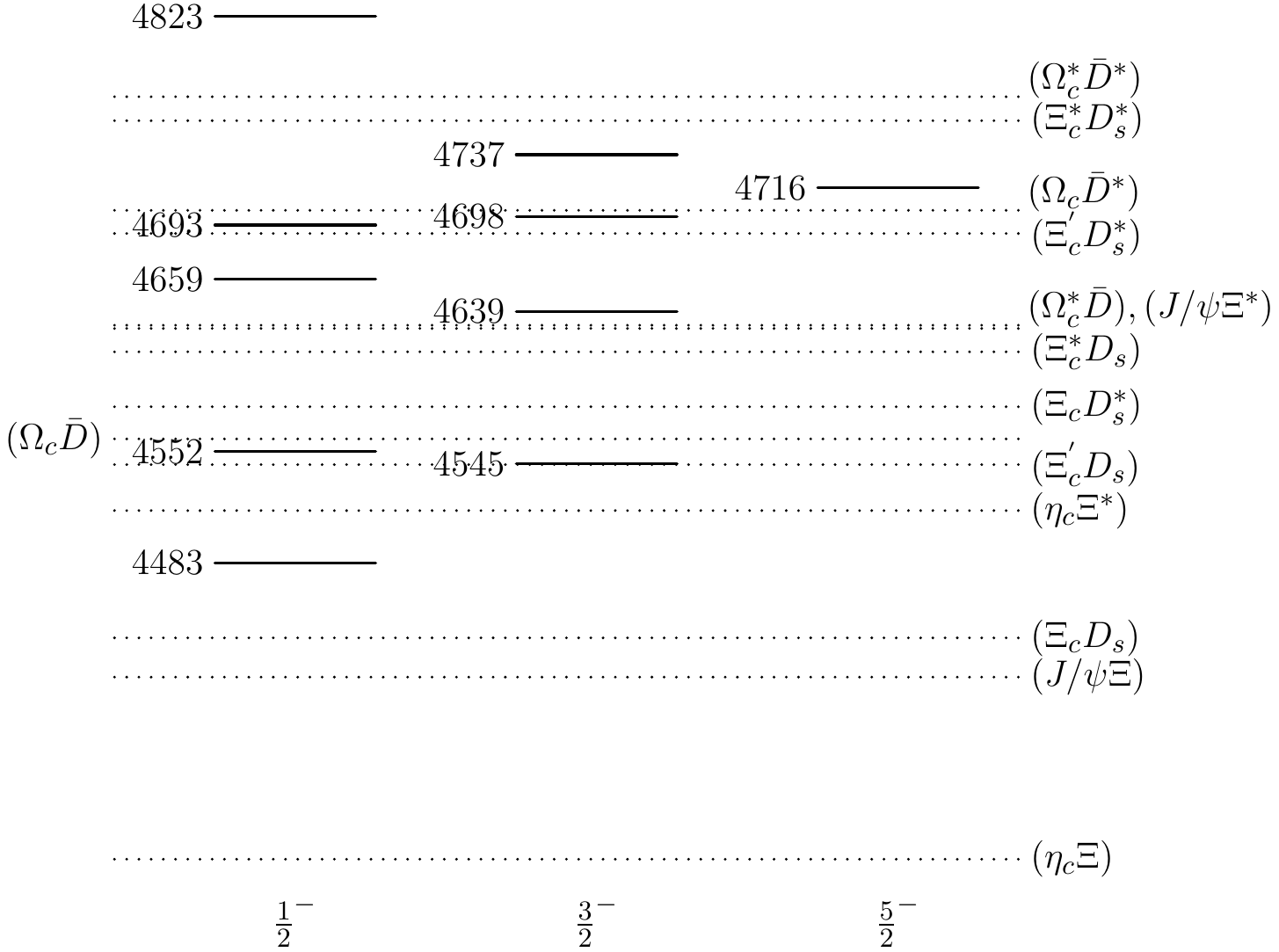}&$\qquad$&
\includegraphics[width=220pt]{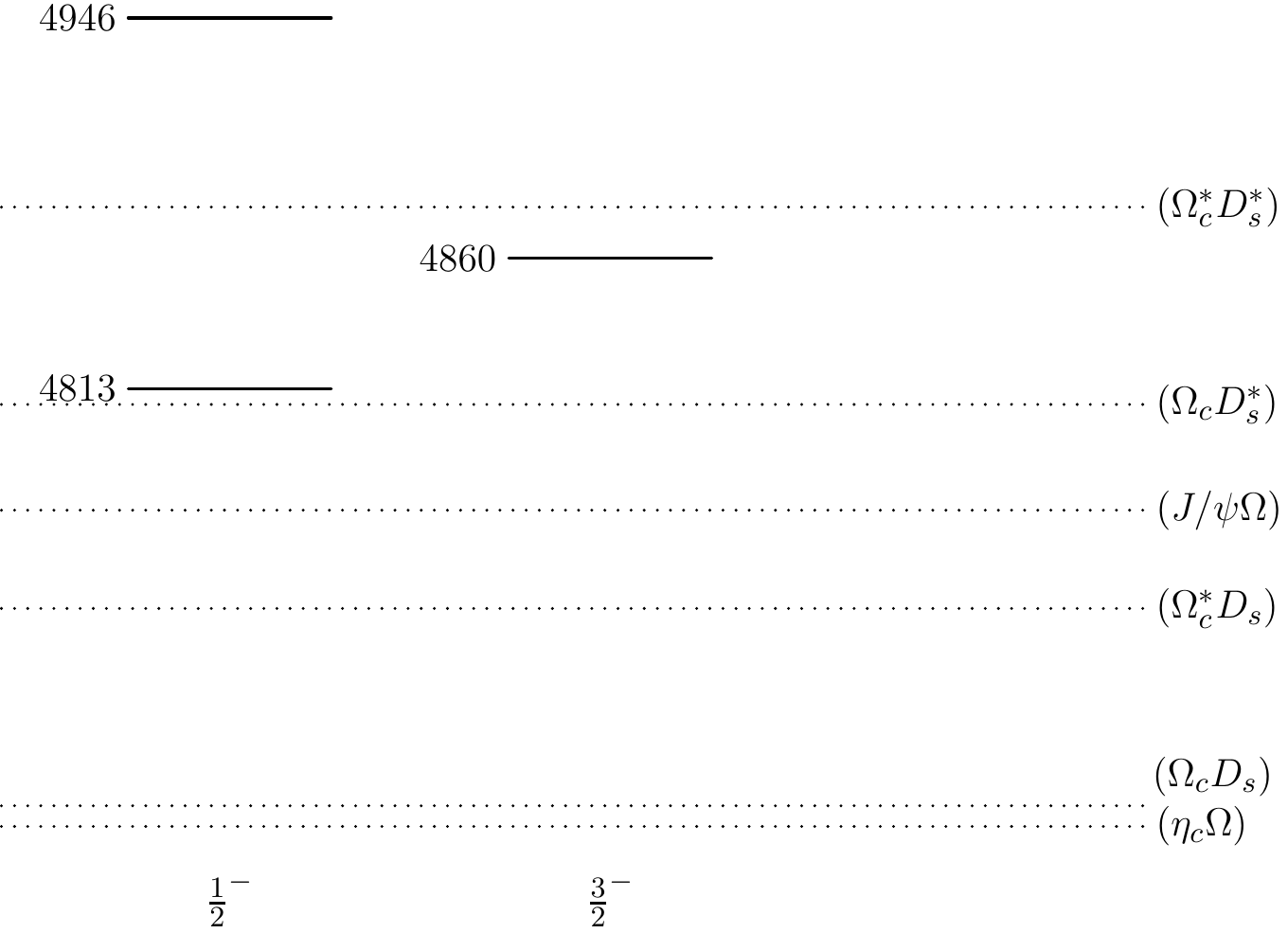}\\
(c) $I=\frac12$ (solid) $ssnc\bar{c}$ states&&(d) $I=0$ (solid)
$sssc\bar{c}$ system
\end{tabular}
\caption{Relative positions for the obtained $qqqc\bar{c}$
pentaquark states. The dotted lines indicate various meson-baryon
thresholds and the long solid line in (a) indicates the observed
isospin-half $P_c$(4380). When a number in the subscript of a
meson-baryon state is equal to the isospin of an initial state, the
decay for the initial state into that meson-baryon channel through
S- or D-wave is allowed. We adopt the masses estimated with the
reference thresholds of $\Sigma_c\bar{D}$ (a), $\Xi_c\bar{D}$ (b),
$\Xi_c D_s$ (c), and $\Omega_cD_s$ (d). The masses are all in units
of MeV.}\label{fig-qqqccbar}
\end{figure}

Fig. \ref{fig-qqqccbar} (a) shows relative positions for these
pentaquarks when one adopts the threshold of $(\Sigma_c\bar{D})$ as
a reference. We also plot all the thresholds of the related
rearrangement decay patterns, i.e. $J/\psi N$, $J/\psi\Delta$,
$\eta_c N$, $\eta_c\Delta$, $\Sigma_c\bar{D}$, $\Sigma_c\bar{D}^*$,
$\Sigma_c^*\bar{D}$, $\Sigma_c^*\bar{D}^*$, $\Lambda_c\bar{D}$, and
$\Lambda_c\bar{D}^*$. The decays may occur through the S- or D-wave
interactions and each pentaquark with $J^P=\frac12^-$, $\frac32^-$,
or $\frac52^-$ can decay to these channels from the parity
conservation and the angular momentum conservation. The isospin
conservation reduces the number of decay channels and we label the
isospin, for convenience, in the subscripts of the meson-baryon
states. Once the considered state is an initial pentaquark plotted
with dashed (solid) line, it can decay into meson-baryon channels
having the subscript $\frac32$ ($\frac12$). Of course, whether the
decay can happen or not is also kinematically constrained by the
pentaquark mass, which depends on models. Contrary to the light
quark case, the decay for the hidden-charm pentaquarks may also get
constraints from the heavy quark symmetry. For the states with
$(I,J)=(\frac32,\frac32)$ and $(I,J)=(\frac12,\frac52)$, the
$c\bar{c}$ spin is always 1. Their decays into $\eta_c\Delta$ and
$\eta_c N$, respectively, involve the heavy quark spin-flip and are
suppressed. In fact, all the hidden-charm decay channels of the
studied pentaquarks are probably suppressed because the transition
from a colored $c\bar{c}$ to a colorless $c\bar{c}$ is a high order
correction of $1/m_c$. With these considerations in mind, it is easy
to judge which channels can be used to search for such unobserved
pentaquark states.

From the results in Fig. \ref{fig-qqqccbar} (a), it is obvious that
the heaviest state is a decuplet baryon with $J=\frac12$. The
lightest state belonging to the flavor octet also has the spin
$J=\frac12$. The observed $P_c(4380)$ is just below the threshold of
$\Sigma_c^*\bar{D}$ and falls in the mass range of the studied
system. Thus, if the estimated masses are reasonable, the
interpretation for the $P_c(4380)$ as a tightly bound pentaquark
with colored $c\bar{c}$ is not excluded, although the present study
cannot give a preferred spin. The partner states decaying into
$J/\psi p$ in the mass region (4280$\sim$4520 MeV) are also
possible. Above the 4520 MeV, the observation of pentaquark-like
baryons in the invariant mass of $J/\psi\Delta$ is possible, too.
All these pentaquarks have open-charm decay channels and are
probably broad states. However, their decays into hidden-charm
channels should have a small fraction. The feature of the broad
width does not contradict with the observed $P_c(4380)$. The
branching ratios for various decay channels will be crucial
information to understand its nature. Note that the decays of the
$J^P=\frac52^-$ state (slightly below the $\Sigma_c^*\bar{D}^*$
threshold) into the channels in the figure are all through $D$ wave.
The width of this state should not be so broad. If our result is
overestimated, both the mass and the width seem not to be
contradicted with those of the $P_c(4450)$, although the parity is
opposite to the preferred $P=+$. In the literature, various
calculations also find a $J^P=\frac52^-$ state below the
$\Sigma_c^*\bar{D}^*$ threshold. If the parity of the $P_c(4450)$ is
really $+$, search for such a negative parity state is also strongly
called for. It is difficult to understand the nature of the
$P_c(4450)$ without further investigations.

Now we discuss the masses estimated with the $J/\psi N$ threshold.
As mentioned in Ref. \cite{Wu:2016gas}, the obtained masses seem to
be underestimated and can be treated as lower limits in this simple
model. The argument is based on the formulae $M=\sum_im_i+\langle
H_{CM}\rangle$ and $M=M_{ref}-\langle H_{CM}\rangle_{ref}+\langle
H_{CM}\rangle$, where the effective quark masses are assumed to be
equal for various hadrons. Of course this assumption leads to
uncertainty for the hadron mass estimation. We illustrate this
uncertainty with the $J/\psi N$ threshold. Because the used
$m_c=1724.8$ MeV gives overestimated $m_{J/\psi}$, it is not
surprising that the former formula results in overestimated masses,
where a charm-anticharm pair exists. When using the latter formula,
in principle, one should have $M_{ref}^{th}=2m_c+3m_n+\langle
H_{CM}\rangle_{ref}$ in order to cancel the quark mass dependence in
the former formula. Because we have adopted
$M_{ref}=m_{J/\psi}+m_N=2m_c^\psi+3m_n^N+\langle
H_{CM}\rangle_{ref}$ but $m_c^\psi<m_c$ and $m_n^N=m_n=361.8$ MeV,
the resulting masses are underestimated. Even in this low mass
limit, the $I=\frac12$ ($I=\frac32$) states above the threshold of
$\Lambda_c\bar{D}$ ($\Sigma_c\bar{D}$) should be broad.

From the above arguments, probably the masses estimated with the
$\Sigma_c\bar{D}$ threshold are more reasonable. If this is the
case, more hidden-charm pentaquarks that can decay into
$J/\psi\Delta$ or $J/\psi p$ are allowed. After a dynamical
calculation with more potential terms is performed in a future work,
one can get more information for the masses of these hidden-charm
pentaquarks. From the obtained masses, various thresholds in Fig.
\ref{fig-qqqccbar} (a), and the above arguments, most pentaquarks
are probably broad states with a small fraction for the hidden-charm
decays. This feature should be different from the molecule picture,
where the hidden-charm decays through rearrangement mechanisms are
probably not suppressed. The exceptional case is for the
$J^P=\frac52^-$ state. Its decays are through $D$ wave and its width
is probably not so broad. If the LHCb hidden-charm pentaquark states
are confirmed, models of rearrangement decays need to be constructed
on one hand, and the measurement of various branching ratios and the
search for more proposed states, on the other hand, are strongly
called for. Of course, the interference effects around the mass
region would make the experimental analysis more difficult.

\subsection{The $nnsc\bar{c}$ system}

\begin{table}[!h]
\caption{Calculated CMI's and estimated pentaquark masses of the
$nnsc\bar{c}$ systems in units of MeV. The masses in the forth
column are calculated with the effective quark masses and are
theoretical upper limits.}\label{mass-nnsccbar}
\begin{tabular}{c|cccccc}\hline
\multicolumn{7}{c}{$nnsc\bar{c}$ ($I=1$)} \\\hline\hline $J^{P}$ &
$\langle H_{CM} \rangle$ &Eigenvalue &Mass&$(J/\psi\Sigma)$
&$(\Sigma_cD_s)$&$(\Xi_c\bar{D})$\\\hline
$\frac52^{-}$ &102.5&102.5&4816.1&4443.8&4625.6&4641.8\\
$\frac32^{-}$ &$\left(\begin{array}{cccc}-6.2&-19.5&35.6&34.0\\-19.5&51.4&34.3&35.6\\35.6&34.3&67.9&-59.2\\34.0&35.6&-59.2&78.4\end{array}\right)$&$\left(\begin{array}{c}132.8\\87.3\\-65.7\\37.2\end{array}\right)$&$\left(\begin{array}{c}4846.4\\4800.9\\4647.9\\4750.8\end{array}\right)$&$\left(\begin{array}{c}4474.0\\4428.5\\4275.5\\4378.4\end{array}\right)$&$\left(\begin{array}{c}4655.9\\4610.4\\4457.4\\4560.3\end{array}\right)$&$\left(\begin{array}{c}4672.0\\4626.5\\4473.5\\4576.5\end{array}\right)$\\
$\frac12^{-}$ &$\left(\begin{array}{ccccc}-71.5&22.5&-48.6&21.5&-50.3\\22.5&50.4&25.6&-104.8&-35.6\\-48.6&25.6&76.2&-35.6&-74.4\\21.5&-104.8&-35.6&53.4&29.9\\-50.3&-35.6&-74.4&29.9&84.2\end{array}\right)$&$\left(\begin{array}{c}219.1\\-120.8\\92.4\\51.9\\-49.9\end{array}\right)$&$\left(\begin{array}{c}4932.7\\4592.8\\4806.0\\4765.5\\4663.7\end{array}\right)$&$\left(\begin{array}{c}4560.3\\4220.5\\4433.6\\4393.1\\4291.3\end{array}\right)$&$\left(\begin{array}{c}4742.2\\4402.3\\4615.5\\4575.0\\4473.2\end{array}\right)$&$\left(\begin{array}{c}4758.4\\4418.5\\4631.6\\4591.1\\4489.3\end{array}\right)$\\
\hline\hline\multicolumn{7}{c}{$nnsc\bar{c}$ ($I=0$)} \\\hline\hline
$J^{P}$ & $\langle H_{CM} \rangle$ &Eigenvalue
&Mass&$(J/\psi\Lambda)$ &$(\Lambda_cD_s)$&$(\Xi_c\bar{D})$\\\hline
$\frac52^{-}$ &79.6&79.6&4793.2&4411.1&4588.8&4618.9\\
$\frac32^{-}$ &$\left(\begin{array}{cccc}-31.0&-18.2&-30.8&34.0\\-18.2&27.4&-38.0&35.6\\-30.8&-38.0&-74.8&-59.2\\34.0&35.6&-59.2&-113.1\end{array}\right)$&$\left(\begin{array}{c}-157.5\\-95.8\\58.2\\3.6\end{array}\right)$&$\left(\begin{array}{c}4556.1\\4617.8\\4771.8\\4717.2\end{array}\right)$&$\left(\begin{array}{c}4174.0\\4235.8\\4389.7\\4335.2\end{array}\right)$&$\left(\begin{array}{c}4351.7\\4413.4\\4567.4\\4512.8\end{array}\right)$&$\left(\begin{array}{c}4381.7\\4443.5\\4597.5\\4542.9\end{array}\right)$\\
$\frac12^{-}$ &$\left(\begin{array}{ccccc}-97.3&-19.5&53.8&21.5&-50.3\\-19.5&-182.5&-46.1&-104.8&-35.6\\53.8&-46.1&-96.6&-35.6&-74.4\\21.5&-104.8&-35.6&-226.1&-43.1\\-50.3&-35.6&-74.4&-43.1&-136.6\end{array}\right)$&$\left(\begin{array}{c}-351.3\\-165.4\\-142.2\\-96.2\\16.0\end{array}\right)$&$\left(\begin{array}{c}4362.3\\4548.2\\4571.4\\4617.4\\4729.6\end{array}\right)$&$\left(\begin{array}{c}3980.3\\4166.1\\4189.4\\4235.4\\4347.5\end{array}\right)$&$\left(\begin{array}{c}4157.9\\4343.8\\4367.0\\4413.0\\4525.2\end{array}\right)$&$\left(\begin{array}{c}4188.0\\4373.8\\4397.1\\4443.1\\4555.2\end{array}\right)$\\
\hline
\end{tabular}
\end{table}

This case is related with the $\Sigma$-like or $\Lambda$-like
baryons with an excited charm-anticharm pair. In Refs.
\cite{Wu:2010jy,Wu:2010vk}, the $\Lambda$-like molecules above 4.2
GeV were predicted. Higher states around 4.6 GeV are also proposed
in Refs. \cite{Chen:2015sxa,Lu:2016roh}. From a calculation with the
one-meson-exchange model \cite{Chen:2016ryt}, several (charmed
baryon)-(charmed strange meson) and (charmed strange
baryon)-(charmed meson) type molecules above 4.5 GeV are possible.
Now we discuss the mass spectrum of a compact structure.

For the compact pentaquarks with colored $c\bar{c}$, the masses of
the isovector (isoscalar) states result from the mixing of the
$10_f$, $8_f(1)$, and $8_f(2)$ ($8_f(1)$, $8_f(2)$, and $1_f$)
induced by the color-magnetic interaction. In the present scheme of
mass estimation, there are three types of reference thresholds:
(charmonium)+($nns$ baryon), ($nnc$ baryon)+($\bar{c}s$ meson), and
($nsc$ baryon)+($\bar{c}n$ meson), which give three sets of masses.
The first threshold gives the lower limits of the masses and the
masses estimated with the other thresholds are slightly different.
We present the numerical results in Tab. \ref{mass-nnsccbar}.

In Fig. \ref{fig-qqqccbar} (b), we show the masses estimated with
the threshold of $\Xi_c\bar{D}$ and various thresholds of relevant
meson-baryon states into which the pentaquarks may decay. The
rearrangement decays involve 18 channels in total. Similar to the
$nnnc\bar{c}$ case, these channels are not forbidden by parity and
angular momentum conservations, and the decays constrained by the
isospin conservation and kinematics are easy to identify from the
figure. As for the constraint from the heavy quark symmetry, the
hidden-charm decay channels are suppressed by the color transition
from octet to singlet. For the decays of the $J=\frac52$ states
(both $I=0$ and $I=0$) into the $\eta_c$ channels, the suppression
is also a heavy quark spin-flip type.

Compared with the $nnnc\bar{c}$ case, the number of pentaquark
masses and that of rearrangement decay patterns are both larger. It
is obvious that most states have open-charm decay channels except
the lightest one with $(I,J)=(0,\frac12)$. If the existence of
open-charm channels means broad widths, the spectrum indicates that
most states are broad and that a narrow pentaquark is possible in
the $nnsc\bar{c}$ system. The dominant channel for this narrow
baryon is $\eta_c\Lambda$. Even if the mass is underestimated about
100 MeV, its open-charm decay channel is still not opened and its
narrow nature does not change. In the extreme case that the mass
approaches its upper limit (4362 MeV), two open-charm channels
$\Xi_c\bar{D}$ and $\Lambda_cD_s$ are opened but it should be still
narrower than other states. The masses estimated with the
$\Sigma_cD_s$ ($\Lambda_cD_s$) threshold are 16 (30) MeV lower than
those with the $\Xi_c\bar{D}$ threshold. Since this number is not
large, the main features do not change. Considering the lower limit
in the present model, we may refine the mass region for this state
to be 3980$\sim$4360 MeV. Note that the predicted $\Lambda$-like
states in the molecule picture in Refs. \cite{Wu:2010jy,Wu:2010vk}
are slightly above this pentaquark. The search for such a state in
the $\eta_c\Lambda$ channel or $J/\psi\Lambda$ channel is strongly
called for.

The CMI in the $SU_f(3)$ symmetric case (Eq. (\ref{cmi10})-Eq.
(\ref{cmi82})) may give us a hint why the low mass and thus narrow
$\Lambda$-type pentaquark is possible. There are two matrix elements
with obviously negative values, one in $1_f$
($S_{c\bar{c}}=1,J=\frac12$) and the other in $8_f(2)$
($S_{c\bar{c}}=1,J=\frac12$). Relevant states can both be isoscalar
spin-half states. Two more states with the same $(I,J)$, one in
$1_f$ ($S_{c\bar{c}}=0,J=\frac12$) and the other in $8_f(2)$
($S_{c\bar{c}}=0,J=\frac12$), have also negative matrix elements
because of the relation $C_{c\bar{c}}<C_{qq}$. The mixing between
these states (and a state in $8_f(1)$) will effectively provide
additional attraction for the lowest pentaquark. Therefore, it is
not surprising that the compact $J^P=\frac12^-$ $\Lambda$-like
pentaquark with colored $c\bar{c}$ has a low mass. Numerically, the
attractive nature is illustrated in Tab. \ref{mass-nnsccbar}, where
the smallest diagonal CMI matrix elements ($-182.5$ MeV, $-226.1$
MeV, and $-136.6$ MeV) all appear in the case $(I,J)=(0,\frac12)$.
The structure mixing effect further induces the lower value $-351.3$
MeV, which corresponds to the lowest $nnsc\bar{c}$ state.

In addition to this narrow compact pentaquark with $J^P=\frac12^-$,
the $J^P=\frac52^-$ $\Lambda$-like state around 4.6 GeV is probably
not a broad one because its dominant open-charm decays are through
$D$-wave. Experimentally, it can be searched for in the
$J/\psi\Lambda$ channel. Since the $\Sigma$-like state with
$J^P=\frac52^-$ is around the threshold of $\Sigma_c^*D_s^*$ and the
present model only gives a rough estimation of its mass, whether it
has a broad width or not needs further study. A possible channel to
search for it is in the $J/\psi\Sigma$ invariant mass distribution.

\subsection{The $ssnc\bar{c}$ and the $sssc\bar{c}$ systems}

\begin{table}[!h]
\caption{Calculated CMI's and estimated pentaquark masses of the
$ssnc\bar{c}$ and $sssc\bar{c}$ systems in units of MeV. The masses
in the forth column are calculated with the effective quark masses
and are theoretical upper limits.}\label{mass-ssnccbar}
\begin{tabular}{c|cccccc}\hline
\multicolumn{7}{c}{$ssnc\bar{c}$ ($I=\frac12$)} \\\hline\hline
$J^{P}$ & $\langle H_{CM} \rangle$ &Eigenvalue &Mass&$(J/\psi\Xi)$
&$(\Omega_c\bar{D})$&$(\Xi_c D_s)$\\\hline
$\frac52^{-}$ &73.5&73.5&4965.7&4571.9&4774.5&4716.2\\
$\frac32^{-}$ &$\left(\begin{array}{cccc}-39.8&-16.1&31.2&32.8\\-16.1&19.7&37.7&36.5\\31.2&37.7&39.0&-59.7\\32.8&36.5&-59.7&28.7\end{array}\right)$&$\left(\begin{array}{c}-98.2\\94.0\\55.4\\-3.6\end{array}\right)$&$\left(\begin{array}{c}4794.0\\4986.2\\4947.6\\4888.6\end{array}\right)$&$\left(\begin{array}{c}4400.2\\4592.5\\4553.9\\4494.9\end{array}\right)$&$\left(\begin{array}{c}4602.8\\4795.1\\4756.5\\4697.5\end{array}\right)$&$\left(\begin{array}{c}4544.5\\4736.8\\4698.2\\4639.2\end{array}\right)$\\
$\frac12^{-}$ &$\left(\begin{array}{ccccc}-107.7&19.7&-53.3&20.7&-51.6\\19.7&13.0&30.5&-103.7&-36.5\\-53.3&30.5&44.5&-36.5&-74.4\\20.7&-103.7&-36.5&10.1&26.2\\-51.6&-36.5&-74.4&26.2&36.6\end{array}\right)$&$\left(\begin{array}{c}180.1\\-159.9\\-90.6\\50.3\\16.5\end{array}\right)$&$\left(\begin{array}{c}5072.3\\4732.3\\4801.6\\4942.5\\4908.7\end{array}\right)$&$\left(\begin{array}{c}4678.5\\4338.6\\4407.9\\4548.7\\4515.0\end{array}\right)$&$\left(\begin{array}{c}4881.1\\4541.2\\4610.5\\4751.3\\4717.6\end{array}\right)$&$\left(\begin{array}{c}4822.8\\4482.9\\4552.2\\4693.0\\4659.3\end{array}\right)$\\
\hline\hline\multicolumn{7}{c}{$sssc\bar{c}$ ($I=0$)} \\\hline\hline
$J^{P}$ & $\langle H_{CM} \rangle$ &Eigenvalue
&Mass&$(J/\psi\Omega)$&$(\Omega_c D_s)$\\\hline
$\frac32^{-}$ &55.1&55.1&5125.9&4744.3&4859.7\\
$\frac12^{-}$ &$\left(\begin{array}{cc}74.1&66.4\\66.4&75.6\end{array}\right)$&$\left(\begin{array}{c}141.3\\8.5\end{array}\right)$&$\left(\begin{array}{c}5212.1\\5079.3\end{array}\right)$&$\left(\begin{array}{c}4830.4\\4697.6\end{array}\right)$&$\left(\begin{array}{c}4945.8\\4813.0\end{array}\right)$\\
\hline
\end{tabular}
\end{table}

Similar to the $\Sigma$-like pentaquarks, all these $\Xi$-like
$ssnc\bar{c}$ states result from the mixing between the $10_f$,
$8_f(1)$, and $8_f(2)$ multiplets. To estimate the masses of these
pentaquark states, we also have three types of reference thresholds
to use: (charmonium)+($ssn$ baryon), ($ssc$ baryon)+($\bar{c}n$
meson), and ($nsc$ baryon)+($\bar{c}s$ meson). The first threshold
gives the lower limits of masses and the other two thresholds result
in two sets of masses with their difference being 60 MeV. The
numerical results are shown in Tab. \ref{mass-ssnccbar}.

To understand the properties further, we plot in Fig.
\ref{fig-qqqccbar} (c) various thresholds and the relative positions
of these states with the masses estimated with the threshold of
$\Xi_cD_s$. Unlike the $nnnc\bar{c}$ and $nnsc\bar{c}$ cases, it is
not necessary to label the isospin in the meson-baryon states since
all the pentaquarks have isospin 1/2. As before, all the given
channels in the figure are not forbidden by parity or angular
momentum conservations but the heavy quark symmetry suppresses the
hidden-charm decays. The remaining constraint comes only from the
kinematics. Probably all these states are broad except the
$J^P=\frac52^-$ one, similar to the $nnnc\bar{c}$ case, because all
the states have open-charm decay channels but the decays of the
$J^P=\frac52^-$ state are through $D$-wave. The estimation with the
threshold of $\Omega_c\bar{D}$ does not change the main decay nature
of these pentaquarks.

The properties of the $\Omega$-like hidden-charm pentaquarks
$sssc\bar{c}$ are similar to those of the $I=\frac32$ $nnnc\bar{c}$
states. We also present the numerical results in Tab.
\ref{mass-ssnccbar}. The relative positions plotted with the masses
estimated with the threshold of $\Omega_cD_s$ and the decay
properties can be found in Fig. \ref{fig-qqqccbar} (d). They should
all be broad states.

\section{Hidden-bottom and $B_c$-like pentaquarks}\label{sec5}

The formulas can be easily applied to much heavier hidden-bottom
pentaquarks. If hidden-charm pentaquarks really exist, their bottom
partners are more likely to form because of the less kinetic energy
in the Hamiltonian. Before the observation of the exotic $P_c$
baryons, such hidden-bottom states had been investigated in Refs.
\cite{Wu:2010rv,Xiao:2013jla,Yang:2011wz}. For recent studies in
different scenarios, one may consult Refs.
\cite{Kopeliovich:2015vqa,Chen:2016qju,Chen:2016heh} for an
overview.

\begin{table}[!h]
\caption{Calculated CMI's and estimated pentaquark masses of the
$nnnb\bar{b}$ systems in units of MeV. The masses in the forth
column are calculated with the effective quark masses and are
theoretical upper limits.}\label{mass-nnnbbbar}
\begin{tabular}{c|ccccc}\hline
\multicolumn{6}{c}{$nnnb\bar{b}$ ($I=\frac32$)} \\\hline\hline
$J^{P}$ & $\langle H_{CM} \rangle$ &Eigenvalue
&Mass&$(\Upsilon\Delta)$ &$(\Sigma_b B)$\\\hline
$\frac32^{-}$ &179.4&179.4&11370.6&10709.6&11268.5\\
$\frac12^{-}$ &$\left(\begin{array}{cc}187.4&19.6\\19.6&189.8\end{array}\right)$&$\left(\begin{array}{c}208.3\\168.9\end{array}\right)$&$\left(\begin{array}{c}11399.5\\11360.1\end{array}\right)$&$\left(\begin{array}{c}10738.5\\10699.2\end{array}\right)$&$\left(\begin{array}{c}11297.4\\11258.0\end{array}\right)$\\
\hline\hline\multicolumn{6}{c}{$nnnb\bar{b}$ ($I=\frac12$)}
\\\hline\hline $J^{P}$ & $\langle H_{CM} \rangle$ &Eigenvalue
&Mass&$(\Upsilon N)$&$(\Sigma_bB)$\\\hline
$\frac52^{-}$ &55.3&55.3&11246.5&10585.6&11144.4\\
$\frac32^{-}$ &$\left(\begin{array}{ccc}21.3&-6.2&15.0\\-6.2&42.6&15.8\\15.0&15.8&-31.9\end{array}\right)$&$\left(\begin{array}{c}46.3\\-39.3\\25.0\end{array}\right)$&$\left(\begin{array}{c}11237.5\\11151.9\\11216.2\end{array}\right)$&$\left(\begin{array}{c}10576.6\\10491.0\\10555.3\end{array}\right)$&$\left(\begin{array}{c}11135.4\\11049.8\\11114.1\end{array}\right)$\\
$\frac12^{-}$ &$\left(\begin{array}{ccc}0.9&-22.4&9.5\\-22.4&-31.0&-2.8\\9.5&-2.8&-52.3\end{array}\right)$&$\left(\begin{array}{c}-54.1\\-42.3\\13.8\end{array}\right)$&$\left(\begin{array}{c}11137.1\\11148.9\\11205.0\end{array}\right)$&$\left(\begin{array}{c}10476.3\\10488.1\\10544.2\end{array}\right)$&$\left(\begin{array}{c}11035.0\\11046.8\\11102.9\end{array}\right)$\\
\hline
\end{tabular}
\end{table}

\begin{table}[!h]
\caption{Calculated CMI's and estimated pentaquark masses of the
$nnsb\bar{b}$ systems in units of MeV. The masses in the forth
column are calculated with the effective quark masses and are
theoretical upper limits.}\label{mass-nnsbbbar}
\begin{tabular}{c|cccccc}\hline
\multicolumn{7}{c}{$nnsb\bar{b}$ ($I=1$)} \\\hline\hline $J^{P}$ &
$\langle H_{CM} \rangle$ &Eigenvalue &Mass&$(\Upsilon\Sigma)$
&$(\Sigma_b{B}_s)$&$(\Xi_b B)$\\\hline
$\frac52^{-}$ &60.1&60.1&11429.9&10777.5&11239.8&11264.1\\
$\frac32^{-}$ &$\left(\begin{array}{cccc}24.7&-7.6&11.8&10.6\\-7.6&46.6&12.4&11.2\\11.8&12.4&65.5&-69.7\\10.6&11.2&-69.7&73.3\end{array}\right)$&$\left(\begin{array}{c}139.2\\52.2\\32.4\\-13.6\end{array}\right)$&$\left(\begin{array}{c}11509.0\\11422.0\\11402.2\\11356.2\end{array}\right)$&$\left(\begin{array}{c}10856.6\\10769.6\\10749.8\\10703.8\end{array}\right)$&$\left(\begin{array}{c}11318.9\\11231.9\\11212.1\\11166.1\end{array}\right)$&$\left(\begin{array}{c}11343.2\\11256.2\\11236.4\\11190.4\end{array}\right)$\\
$\frac12^{-}$ &$\left(\begin{array}{ccccc}3.5&7.4&-17.6&6.7&-15.8\\7.4&60.1&9.0&-83.9&-11.2\\-17.6&9.0&71.4&-11.2&-74.4\\6.7&-83.9&-11.2&68.3&8.2\\-15.8&-11.2&-74.4&8.2&79.4\end{array}\right)$&$\left(\begin{array}{c}168.8\\129.2\\27.6\\-26.4\\-16.6\end{array}\right)$&$\left(\begin{array}{c}11538.6\\11499.0\\11397.4\\11343.4\\11353.2\end{array}\right)$&$\left(\begin{array}{c}10886.3\\10846.7\\10745.1\\10691.0\\10700.9\end{array}\right)$&$\left(\begin{array}{c}11348.5\\11308.9\\11207.3\\11153.3\\11163.1\end{array}\right)$&$\left(\begin{array}{c}11372.9\\11333.2\\11231.6\\11177.6\\11187.4\end{array}\right)$\\
\hline\hline\multicolumn{7}{c}{$nnsb\bar{b}$ ($I=0$)} \\\hline\hline
$J^{P}$ & $\langle H_{CM} \rangle$ &Eigenvalue
&Mass&$(\Upsilon\Lambda)$ &$(\Lambda_b{B}_s)$&$(\Xi_b B)$\\\hline
$\frac52^{-}$ &34.9&34.9&11404.7&10742.6&11205.2&11238.9\\
$\frac32^{-}$ &$\left(\begin{array}{cccc}1.5&-6.4&-11.2&10.6\\-6.4&22.6&-11.0&11.2\\-11.2&-11.0&-97.5&-69.7\\10.6&11.2&-69.7&-136.8\end{array}\right)$&$\left(\begin{array}{c}-189.6\\-53.0\\26.4\\6.0\end{array}\right)$&$\left(\begin{array}{c}11180.2\\11316.8\\11396.2\\11375.8\end{array}\right)$&$\left(\begin{array}{c}10518.2\\10654.7\\10734.2\\10713.8\end{array}\right)$&$\left(\begin{array}{c}10980.7\\11117.3\\11196.7\\11176.3\end{array}\right)$&$\left(\begin{array}{c}11014.4\\11151.0\\11230.4\\11210.0\end{array}\right)$\\
$\frac12^{-}$ &$\left(\begin{array}{ccccc}-18.5&-7.1&15.6&6.7&-15.8\\-7.1&-132.5&-13.9&-83.9&-11.2\\15.6&-13.9&-101.4&-11.2&-74.4\\6.7&-83.9&-11.2&-173.8&-15.8\\-15.8&-11.2&-74.4&-15.8&-141.4\end{array}\right)$&$\left(\begin{array}{c}-252.1\\-186.0\\-68.6\\-56.7\\-4.1\end{array}\right)$&$\left(\begin{array}{c}11117.7\\11183.8\\11301.2\\11313.1\\11365.7\end{array}\right)$&$\left(\begin{array}{c}10455.7\\10521.7\\10639.1\\10651.0\\10703.6\end{array}\right)$&$\left(\begin{array}{c}10918.2\\10984.3\\11101.7\\11113.6\\11166.2\end{array}\right)$&$\left(\begin{array}{c}10951.9\\11018.0\\11135.4\\11147.3\\11199.9\end{array}\right)$\\
\hline
\end{tabular}
\end{table}

\begin{table}[!h]
\caption{Calculated CMI's and estimated pentaquark masses of the
$ssnb\bar{b}$ and $sssb\bar{b}$ systems in units of MeV. The masses
in the forth column are calculated with the effective quark masses
and are theoretical upper limits.}\label{mass-ssnbbbar}
\begin{tabular}{c|cccccc}\hline
\multicolumn{7}{c}{$ssnb\bar{b}$ ($I=\frac12$)} \\\hline\hline
$J^{P}$ & $\langle H_{CM} \rangle$ &Eigenvalue &Mass&$(\Upsilon\Xi)$
&$(\Omega_bB)$&$(\Xi_b{B}_s)$\\\hline
$\frac52^{-}$ &27.3&27.3&11575.7&10902.0&11383.8&11321.9\\
$\frac32^{-}$ &$\left(\begin{array}{cccc}-6.3&-7.1&10.7&11.9\\-7.1&14.9&10.8&12.0\\10.7&10.8&33.9&-69.1\\11.9&12.0&-69.1&26.1\end{array}\right)$&$\left(\begin{array}{c}99.2\\-50.5\\19.6\\0.2\end{array}\right)$&$\left(\begin{array}{c}11647.6\\11497.9\\11568.0\\11548.6\end{array}\right)$&$\left(\begin{array}{c}10973.8\\10824.2\\10894.3\\10874.9\end{array}\right)$&$\left(\begin{array}{c}11455.6\\11306.0\\11376.1\\11356.7\end{array}\right)$&$\left(\begin{array}{c}11393.8\\11244.1\\11314.2\\11294.8\end{array}\right)$\\
$\frac12^{-}$ &$\left(\begin{array}{ccccc}-26.5&6.8&-15.2&7.5&-17.0\\6.8&28.1&7.6&-85.1&-12.0\\-15.2&7.6&39.7&-12.0&-74.4\\7.5&-85.1&-12.0&19.9&8.4\\-17.0&-12.0&-74.4&8.4&31.8\end{array}\right)$&$\left(\begin{array}{c}129.7\\89.7\\-64.3\\-55.2\\-7.0\end{array}\right)$&$\left(\begin{array}{c}11678.1\\11638.1\\11484.1\\11493.2\\11541.4\end{array}\right)$&$\left(\begin{array}{c}11004.4\\10964.4\\10810.4\\10819.5\\10867.7\end{array}\right)$&$\left(\begin{array}{c}11486.2\\11446.2\\11292.2\\11301.3\\11349.5\end{array}\right)$&$\left(\begin{array}{c}11424.3\\11384.3\\11230.3\\11239.4\\11287.6\end{array}\right)$\\
\hline\hline\multicolumn{7}{c}{$sssb\bar{b}$ ($I=0$)} \\\hline\hline
$J^{P}$ & $\langle H_{CM} \rangle$ &Eigenvalue
&Mass&$(\Upsilon\Omega)$&$(\Omega_b{B}_s)$\\\hline
$\frac32^{-}$ &59.4&59.4&11786.4&11124.7&11506.5\\
$\frac12^{-}$ &$\left(\begin{array}{cc}70.4&20.2\\20.2&70.8\end{array}\right)$&$\left(\begin{array}{c}90.8\\50.4\end{array}\right)$&$\left(\begin{array}{c}11817.8\\11777.4\end{array}\right)$&$\left(\begin{array}{c}11156.1\\11115.7\end{array}\right)$&$\left(\begin{array}{c}11537.9\\11497.5\end{array}\right)$\\
\hline
\end{tabular}
\end{table}

\begin{figure}[!h]
\begin{tabular}{ccc}
\includegraphics[width=220pt]{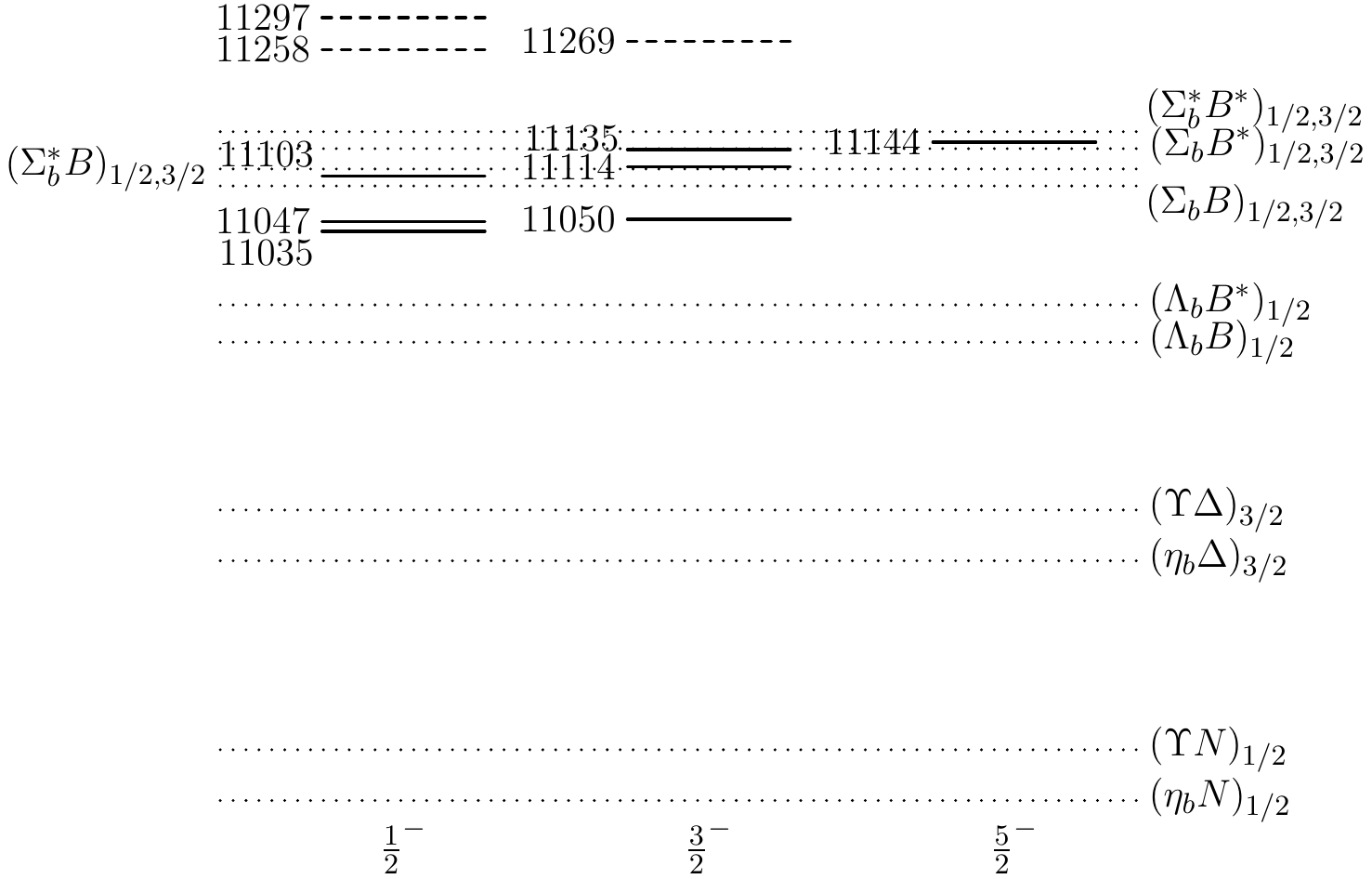}&$\qquad$&
\includegraphics[width=220pt]{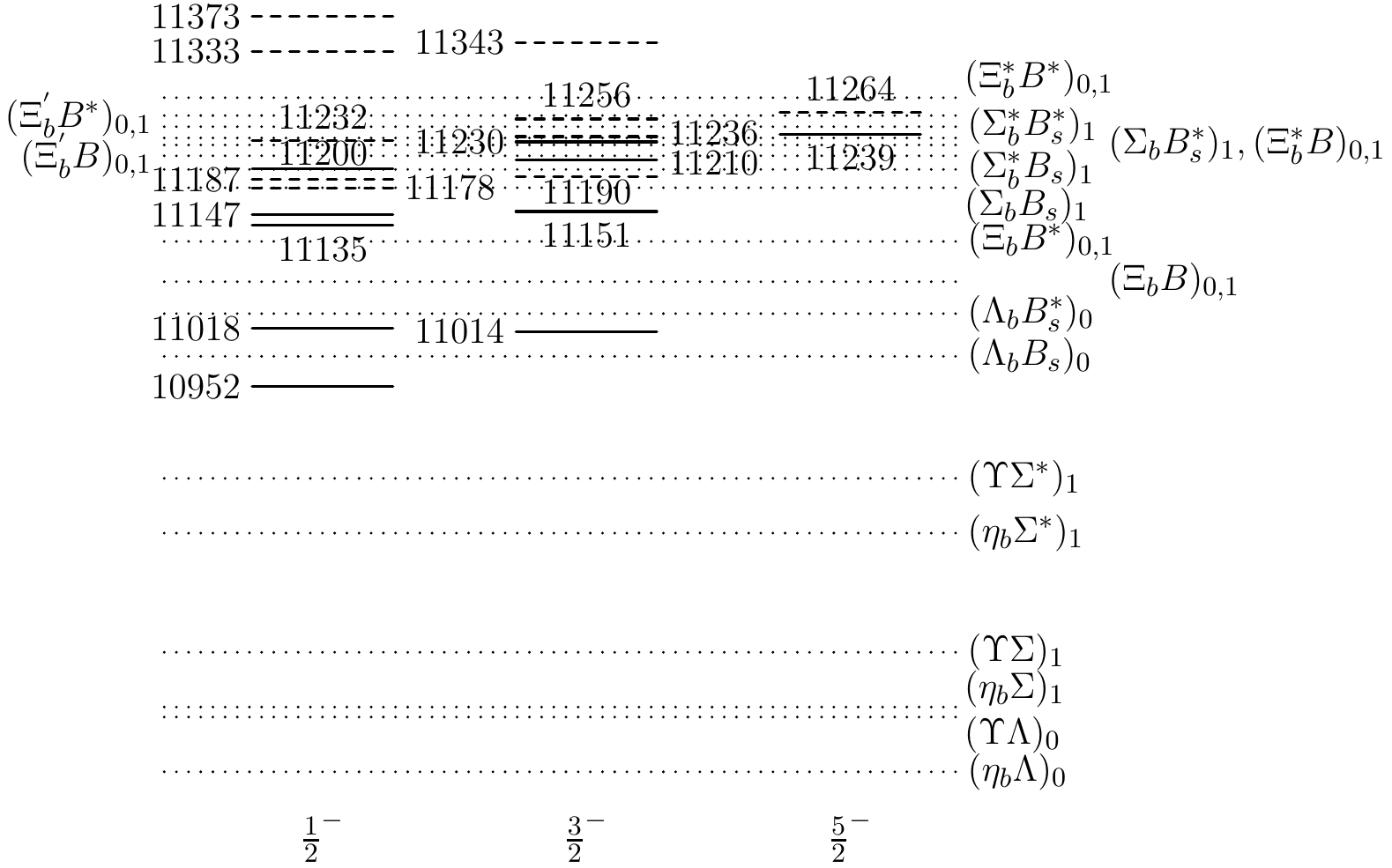}\\
(a) $I=\frac32$ (dashed) and $I=\frac12$ (solid) $nnnb\bar{b}$ states &&(b) $I=1$ (dashed) and $I=0$ (solid) $nnsb\bar{b}$ states\\
&&\\
\includegraphics[width=220pt]{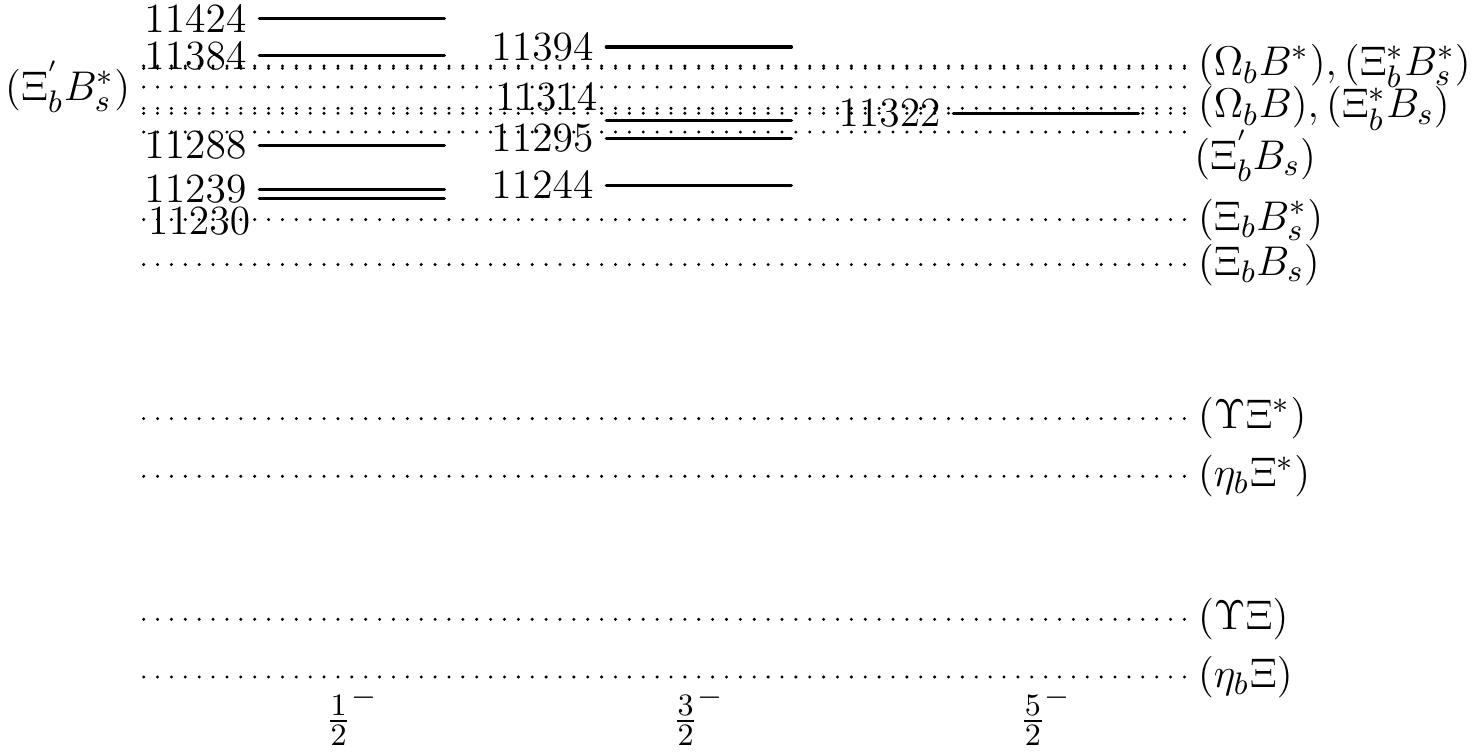}&$\qquad$&
\includegraphics[width=220pt]{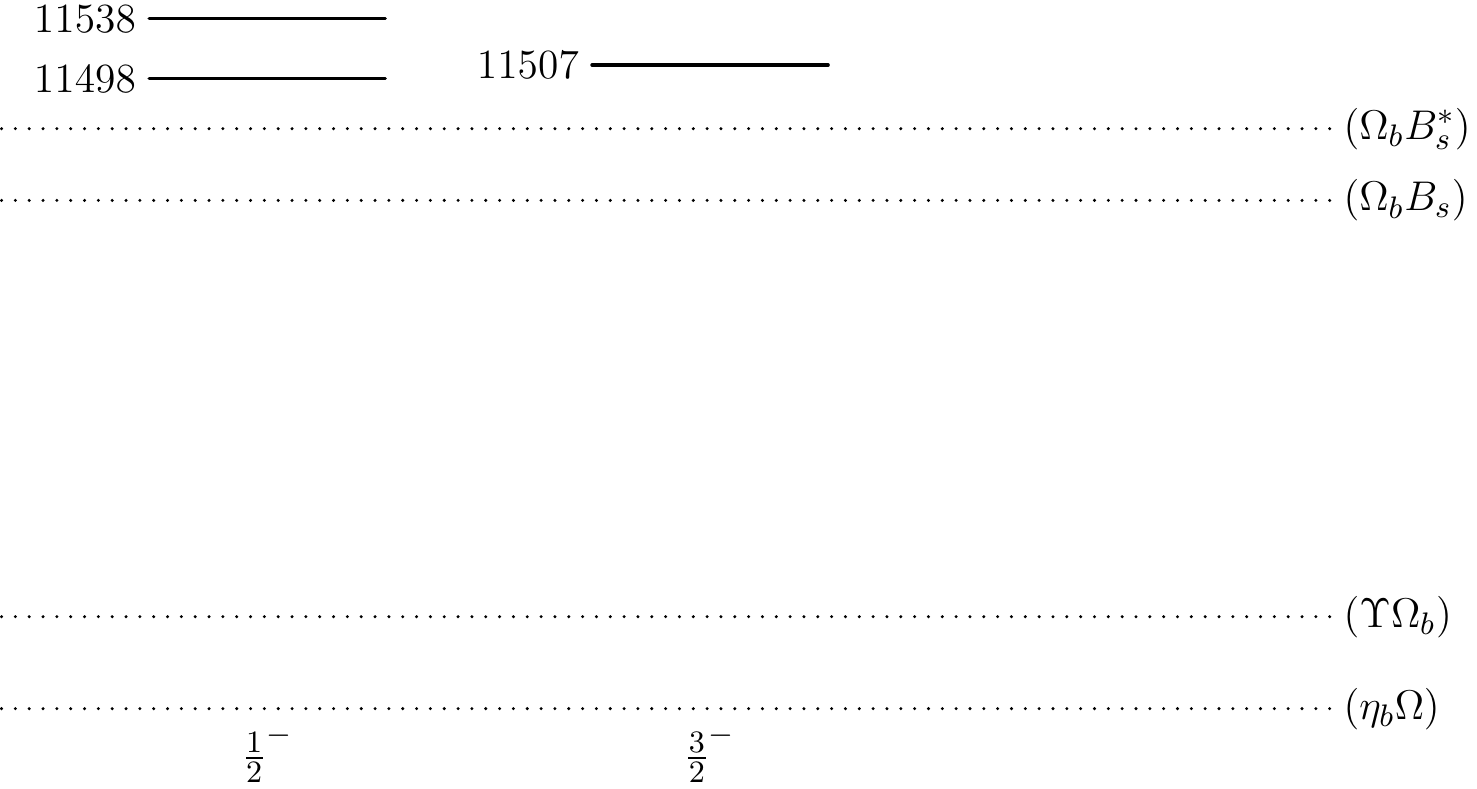}\\
(c) $I=\frac12$ (solid) $ssnb\bar{b}$ states &&(d) $I=0$ (solid) $sssb\bar{b}$ states\\
\end{tabular}
\caption{Relative positions for the obtained $qqqb\bar{b}$
pentaquark states. The dotted lines indicate various meson-baryon
thresholds. When a number in the subscript of a meson-baryon state
is equal to the isospin of an initial state, the decay for the
initial state into that meson-baryon channel through S- or D-wave is
allowed. We adopt the masses estimated with the reference thresholds
of $\Sigma_bB$ (a), $\Xi_bB$ (b), $\Xi_b B_s$ (c), and $\Omega_bB_s$
(d). The masses are all in units of MeV.}\label{fig-qqqbbbar}
\end{figure}

When estimating the hidden-bottom pentaquark masses, the reference
thresholds we use are $\Upsilon\Delta$, $\Upsilon N$, and
$\Sigma_bB$ for the $nnnb\bar{b}$ states, $\Upsilon\Sigma$,
$\Upsilon\Lambda$, $\Sigma_b{B}_s$, $\Lambda_b{B}_s$, and $\Xi_bB$
for the $nnsb\bar{b}$ states, $\Upsilon\Xi$, $\Omega_bB$, and
$\Xi_b{B}_s$ for the $ssnb\bar{b}$ states, and $\Upsilon\Omega$ and
$\Omega_b{B}_s$ for the $sssb\bar{b}$ states. We present estimations
for the masses of these hidden-bottom pentaquarks in Tabs.
\ref{mass-nnnbbbar}, \ref{mass-nnsbbbar}, and \ref{mass-ssnbbbar},
respectively. The relative positions for the $nnnb\bar{b}$,
$nnsb\bar{b}$, $ssnb\bar{b}$, and $sssb\bar{b}$ states are shown in
Fig. \ref{fig-qqqbbbar} with the masses estimated from the
thresholds of $\Sigma_bB$, $\Xi_bB$, $\Xi_bB_s$, and $\Omega_bB_s$,
respectively. The following discussions are based on the assumption
that these estimated masses are reasonable.

The basic features for the mass spectrum and the decay properties of
the $nnnb\bar{b}$ system are similar to the hidden charm case: i)
All the non-strange pentaquarks can decay into $\Upsilon$ (or
$\eta_b$) plus $p$ (or $\Delta$) and all of them have also
open-bottom decay channels; ii) Maybe the width for the $J^P=5/2^-$
state is not so broad while others have broad widths; iii) The
hidden bottom decay channels should have smaller branching ratios
than the molecules do. If we compare results with those in Refs.
\cite{Wu:2010rv,Xiao:2013jla,Yang:2011wz}, where the S-wave
$\Sigma_b\bar{B}$ hadronic molecule was proposed, one can conclude
here that lower hidden-bottom pentaquarks than the $\Sigma_b\bar{B}$
threshold are possible, too.

The $\Sigma$-like or $\Lambda$-like hidden bottom system is more
interesting than the hidden charm case. Now, the states concentrated
around the threshold of $\Sigma_bB_s$ are more than those around
$\Sigma_cD_s$. There are probably three narrow $\Lambda$-like states
with $J^P=\frac12^-$ or $\frac32^-$ and two $J^P=\frac52^-$ states
having relatively narrow widths, which is sensitive to the mass
values. If the states shown in Fig. \ref{fig-qqqbbbar}(b) are
underestimated, say 60 MeV, all the states would have open-bottom
decay channels and broader widths are expected. If they are
overestimated, say 60 MeV, the lowest three states will not have
open-bottom decay channels and should have narrow widths. The S-wave
decay channels $\Sigma_b^*B_s^*$ and $\Xi_b^*B^*$ are both closed
for the $J^P=\frac52^-$ states, too. Anyway, the $\Lambda$-like
states below 11100 MeV are worthwhile study
\cite{Wu:2010rv,Xiao:2013jla}. The searching for such states in the
$\Upsilon\Lambda$ channel will give more information. The study for
the isovector pentaquarks in the $\Upsilon\Sigma$ channel is also
proposed.

For the $\Xi$-like and $\Omega$-like hidden-bottom pentaquark
systems, there is only one candidate having relatively narrow width,
the $J^P=\frac52^-$ $ssnb\bar{b}$ state. If its mass is below the
threshold of $\Xi_b^*B_s^*$, all the two-body open-bottom decays
are through $D$-wave. Although the decay into $\Upsilon\Xi^*$
may be through $S$-wave, its contribution should be suppressed.
One may search for this state in the $\Upsilon\Xi$ channel.\\

\begin{table}[!h]
\caption{Calculated CMI's and estimated pentaquark masses of the
$nnnb\bar{c}$ systems in units of MeV. The masses in the forth
column are calculated with the effective quark masses and are
theoretical upper limits.}\label{mass-nnnbcbar}
\begin{tabular}{c|ccccc}\hline
\multicolumn{6}{c}{$nnnb\bar{c}$ ($I=\frac32$)} \\\hline\hline
$J^{P}$ & $\langle H_{CM} \rangle$ &Eigenvalue &Mass&$(B_c\Delta)$
&$(\Sigma_b \bar{D})$\\\hline
$\frac32^{-}$ &163.8&163.8&8026.9&7577.1&7911.9\\
$\frac12^{-}$ &$\left(\begin{array}{cc}217.8&46.2\\46.2&190.6\end{array}\right)$&$\left(\begin{array}{c}252.3\\156.1\end{array}\right)$&$\left(\begin{array}{c}8115.4\\8019.2\end{array}\right)$&$\left(\begin{array}{c}7665.6\\7569.4\end{array}\right)$&$\left(\begin{array}{c}8000.4\\7904.2\end{array}\right)$\\
\hline\hline\multicolumn{6}{c}{$nnnb\bar{c}$ ($I=\frac12$)}
\\\hline\hline $J^{P}$ & $\langle H_{CM} \rangle$ &Eigenvalue
&Mass&$(B_c N)$&$(\Sigma_b\bar{D})$\\\hline
$\frac52^{-}$ &82.6&82.6&7945.7&7496.0&7830.7\\
$\frac32^{-}$ &$\left(\begin{array}{ccc}2.6&-41.8&53.8\\-41.8&43.4&45.9\\53.8&45.9&-23.0\end{array}\right)$&$\left(\begin{array}{c}-93.5\\72.0\\44.5\end{array}\right)$&$\left(\begin{array}{c}7769.6\\7935.1\\7907.6\end{array}\right)$&$\left(\begin{array}{c}7319.9\\7485.4\\7457.9\end{array}\right)$&$\left(\begin{array}{c}7654.6\\7820.1\\7792.6\end{array}\right)$\\
$\frac12^{-}$ &$\left(\begin{array}{ccc}-45.4&-64.9&34.0\\-64.9&-30.2&-18.7\\34.0&-18.7&-71.0\end{array}\right)$&$\left(\begin{array}{c}-109.1\\-77.3\\39.7\end{array}\right)$&$\left(\begin{array}{c}7754.0\\7785.8\\7902.8\end{array}\right)$&$\left(\begin{array}{c}7304.3\\7336.1\\7453.1\end{array}\right)$&$\left(\begin{array}{c}7639.0\\7670.8\\7787.8\end{array}\right)$\\
\hline
\end{tabular}
\end{table}

\begin{table}[!h]
\caption{Calculated CMI's and estimated pentaquark masses of the
$nnsb\bar{c}$ systems in units of MeV. The masses in the forth
column are calculated with the effective quark masses and are
theoretical upper limits.}\label{mass-nnsbcbar}
\begin{tabular}{c|cccccc}\hline
\multicolumn{7}{c}{$nnsb\bar{c}$ ($I=1$)} \\\hline\hline $J^{P}$ &
$\langle H_{CM} \rangle$ &Eigenvalue &Mass&$(B_c\Sigma)$
&$(\Sigma_bD_s)$&$(\Xi_b\bar{D})$\\\hline
$\frac52^{-}$ &86.5&86.5&8128.2&7687.0&7938.1&7949.5\\
$\frac32^{-}$ &$\left(\begin{array}{cccc}6.7&-42.0&37.8&38.0\\-42.0&47.4&32.6&32.4\\37.8&32.6&62.4&-57.4\\38.0&32.4&-57.4&70.1\end{array}\right)$&$\left(\begin{array}{c}123.9\\79.0\\-77.0\\60.8\end{array}\right)$&$\left(\begin{array}{c}8165.6\\8120.7\\7964.7\\8102.5\end{array}\right)$&$\left(\begin{array}{c}7724.4\\7679.5\\7523.5\\7661.3\end{array}\right)$&$\left(\begin{array}{c}7975.5\\7930.6\\7774.6\\7912.4\end{array}\right)$&$\left(\begin{array}{c}7986.9\\7942.0\\7786.0\\7923.8\end{array}\right)$\\
$\frac12^{-}$ &$\left(\begin{array}{ccccc}-41.2&23.9&-46.1&24.0&-45.9\\23.9&65.3&13.8&-108.4&-32.4\\-46.1&13.8&72.2&-32.4&-74.4\\24.0&-108.4&-32.4&73.9&13.3\\-45.9&-32.4&-74.4&13.3&80.2\end{array}\right)$&$\left(\begin{array}{c}212.4\\116.6\\-94.0\\61.2\\-45.6\end{array}\right)$&$\left(\begin{array}{c}8254.1\\8158.3\\7947.7\\8102.9\\7996.1\end{array}\right)$&$\left(\begin{array}{c}7812.9\\7717.1\\7506.5\\7661.7\\7554.9\end{array}\right)$&$\left(\begin{array}{c}8064.0\\7968.2\\7757.6\\7912.8\\7806.0\end{array}\right)$&$\left(\begin{array}{c}8075.4\\7979.6\\7769.0\\7924.2\\7817.4\end{array}\right)$\\
\hline\hline\multicolumn{7}{c}{$nnsb\bar{c}$ ($I=0$)} \\\hline\hline
$J^{P}$ & $\langle H_{CM} \rangle$ &Eigenvalue &Mass&$(B_c\Lambda)$
&$(\Lambda_bD_s)$&$(\Xi_b\bar{D})$\\\hline
$\frac52^{-}$ &62.3&62.3&8104.0&7653.1&7904.5&7925.3\\
$\frac32^{-}$ &$\left(\begin{array}{cccc}-17.2&-42.2&-38.4&38.0\\-42.2&23.4&-32.1&32.4\\-38.4&-32.1&-76.3&-57.4\\38.0&32.4&-57.4&-116.5\end{array}\right)$&$\left(\begin{array}{c}-159.0\\-107.7\\52.4\\27.6\end{array}\right)$&$\left(\begin{array}{c}7882.7\\7934.0\\8094.1\\8069.3\end{array}\right)$&$\left(\begin{array}{c}7431.8\\7483.1\\7643.2\\7618.4\end{array}\right)$&$\left(\begin{array}{c}7683.2\\7734.5\\7894.6\\7869.8\end{array}\right)$&$\left(\begin{array}{c}7704.0\\7755.3\\7915.4\\7890.6\end{array}\right)$\\
$\frac12^{-}$ &$\left(\begin{array}{ccccc}-65.0&-24.3&45.5&24.0&-45.9\\-24.3&-175.6&-51.0&-108.4&-32.4\\45.5&-51.0&-100.6&-32.4&-74.4\\24.0&-108.4&-32.4&-215.1&-51.4\\-45.9&-32.4&-74.4&-51.4&-140.6\end{array}\right)$&$\left(\begin{array}{c}-350.7\\-156.5\\-121.4\\-91.7\\23.4\end{array}\right)$&$\left(\begin{array}{c}7691.0\\7885.2\\7920.3\\7950.0\\8065.1\end{array}\right)$&$\left(\begin{array}{c}7240.1\\7434.3\\7469.4\\7499.1\\7614.2\end{array}\right)$&$\left(\begin{array}{c}7491.5\\7685.7\\7720.8\\7750.5\\7865.6\end{array}\right)$&$\left(\begin{array}{c}7512.3\\7706.5\\7741.6\\7771.3\\7886.4\end{array}\right)$\\
\hline
\end{tabular}
\end{table}

\begin{table}[!h]
\caption{Calculated CMI's and estimated pentaquark masses of the
$ssnb\bar{c}$ and $sssb\bar{c}$ systems in units of MeV. The masses
in the forth column are calculated with the effective quark masses
and are theoretical upper limits.}\label{mass-ssnbcbar}
\begin{tabular}{c|cccccc}\hline
\multicolumn{7}{c}{$ssnb\bar{c}$ ($I=\frac12$)} \\\hline\hline
$J^{P}$ & $\langle H_{CM} \rangle$ &Eigenvalue &Mass&$(B_c\Xi)$
&$(\Omega_b\bar{D})$&$(\Xi_bD_s)$\\\hline
$\frac52^{-}$ &54.4&54.4&8274.7&7812.1&8069.9&8020.9\\
$\frac32^{-}$ &$\left(\begin{array}{cccc}-24.8&-42.4&38.4&38.2\\-42.4&15.7&32.2&32.3\\38.4&32.2&30.4&-57.3\\38.2&32.3&-57.3&22.8\end{array}\right)$&$\left(\begin{array}{c}-112.3\\84.1\\46.0\\26.3\end{array}\right)$&$\left(\begin{array}{c}8108.0\\8304.4\\8266.3\\8246.6\end{array}\right)$&$\left(\begin{array}{c}7645.4\\7841.8\\7803.7\\7784.0\end{array}\right)$&$\left(\begin{array}{c}7903.1\\8099.6\\8061.4\\8041.7\end{array}\right)$&$\left(\begin{array}{c}7854.2\\8050.6\\8012.5\\7992.8\end{array}\right)$\\
$\frac12^{-}$ &$\left(\begin{array}{ccccc}-72.4&24.3&-45.5&24.1&-45.7\\24.3&34.3&13.2&-108.5&-32.3\\-45.5&13.2&40.5&-32.3&-74.4\\24.1&-108.5&-32.3&25.8&13.7\\-45.7&-32.3&-74.4&13.7&32.6\end{array}\right)$&$\left(\begin{array}{c}172.7\\-128.0\\-85.1\\77.2\\24.1\end{array}\right)$&$\left(\begin{array}{c}8393.0\\8092.3\\8135.2\\8297.5\\8244.4\end{array}\right)$&$\left(\begin{array}{c}7930.4\\7629.7\\7672.6\\7834.9\\7781.9\end{array}\right)$&$\left(\begin{array}{c}8188.1\\7887.5\\7930.3\\8092.6\\8039.6\end{array}\right)$&$\left(\begin{array}{c}8139.2\\7838.5\\7881.4\\8043.7\\7990.6\end{array}\right)$\\
\hline\hline\multicolumn{7}{c}{$sssb\bar{c}$ ($I=0$)} \\\hline\hline
$J^{P}$ & $\langle H_{CM} \rangle$ &Eigenvalue
&Mass&$(B_c\Omega)$&$(\Omega_bD_s)$\\\hline
$\frac32^{-}$ &44.5&44.5&8443.4&7992.9&8163.4\\
$\frac12^{-}$ &$\left(\begin{array}{cc}99.5&45.6\\45.6&71.6\end{array}\right)$&$\left(\begin{array}{c}133.2\\37.8\end{array}\right)$&$\left(\begin{array}{c}8532.1\\8436.7\end{array}\right)$&$\left(\begin{array}{c}8081.6\\7986.2\end{array}\right)$&$\left(\begin{array}{c}8252.2\\8156.8\end{array}\right)$\\
\hline
\end{tabular}
\end{table}

\begin{table}[!h]
\caption{Calculated CMI's and estimated pentaquark masses of the
$nnnc\bar{b}$ systems in units of MeV. The masses in the forth
column are calculated with the effective quark masses and are
theoretical upper limits.}\label{mass-nnncbbar}
\begin{tabular}{c|ccccc}\hline
\multicolumn{6}{c}{$nnnc\bar{b}$ ($I=\frac32$)} \\\hline\hline
$J^{P}$ & $\langle H_{CM} \rangle$ &Eigenvalue &Mass&$(B_c \Delta)$
&$(\Sigma_cB)$\\\hline
$\frac32^{-}$ &188.1&188.1&8051.2&7601.4&7948.7\\
$\frac12^{-}$ &$\left(\begin{array}{cc}169.1&35.2\\35.2&190.6\end{array}\right)$&$\left(\begin{array}{c}216.7\\143.0\end{array}\right)$&$\left(\begin{array}{c}8079.8\\8006.1\end{array}\right)$&$\left(\begin{array}{c}7630.0\\7556.3\end{array}\right)$&$\left(\begin{array}{c}7977.3\\7903.6\end{array}\right)$\\
\hline\hline\multicolumn{6}{c}{$nnnc\bar{b}$ ($I=\frac12$)}
\\\hline\hline $J^{P}$ & $\langle H_{CM} \rangle$ &Eigenvalue
&Mass&$(B_c N)$&$(\Sigma_cB)$\\\hline
$\frac52^{-}$ &71.2&71.2&7934.3&7484.6&7831.8\\
$\frac32^{-}$ &$\left(\begin{array}{ccc}10.2&14.7&9.3\\14.7&43.4&20.2\\9.3&20.2&-26.8\end{array}\right)$&$\left(\begin{array}{c}55.0\\-32.9\\4.7\end{array}\right)$&$\left(\begin{array}{c}7918.1\\7830.2\\7867.8\end{array}\right)$&$\left(\begin{array}{c}7468.4\\7380.5\\7418.1\end{array}\right)$&$\left(\begin{array}{c}7815.6\\7727.7\\7765.3\end{array}\right)$\\
$\frac12^{-}$ &$\left(\begin{array}{ccc}-26.4&-28.6&5.9\\-28.6&-30.2&6.6\\5.9&6.6&-63.4\end{array}\right)$&$\left(\begin{array}{c}-69.6\\-50.8\\0.4\end{array}\right)$&$\left(\begin{array}{c}7793.5\\7812.3\\7863.5\end{array}\right)$&$\left(\begin{array}{c}7343.8\\7362.6\\7413.8\end{array}\right)$&$\left(\begin{array}{c}7691.0\\7709.8\\7761.0\end{array}\right)$\\
\hline
\end{tabular}
\end{table}

\begin{table}[!h]
\caption{Calculated CMI's and estimated pentaquark masses of the
$nnsc\bar{b}$ systems in units of MeV. The masses in the forth
column are calculated with the effective quark masses and are
theoretical upper limits.}\label{mass-nnscbbar}
\begin{tabular}{c|cccccc}\hline
\multicolumn{7}{c}{$nnsc\bar{b}$ ($I=1$)} \\\hline\hline $J^{P}$ &
$\langle H_{CM} \rangle$ &Eigenvalue &Mass&$(B_c\Sigma)$
&$(\Sigma_cB_s)$&$(\Xi_cB)$\\\hline
$\frac52^{-}$ &77.2&77.2&8118.9&7677.7&7928.4&7957.5\\
$\frac32^{-}$ &$\left(\begin{array}{cccc}12.9&14.9&9.5&6.6\\14.9&47.4&14.2&14.3\\9.5&14.2&72.0&-71.5\\6.6&14.3&-71.5&82.7\end{array}\right)$&$\left(\begin{array}{c}149.0\\62.2\\8.3\\-4.6\end{array}\right)$&$\left(\begin{array}{c}8190.7\\8103.9\\8050.0\\8037.1\end{array}\right)$&$\left(\begin{array}{c}7749.5\\7662.7\\7608.8\\7595.9\end{array}\right)$&$\left(\begin{array}{c}8000.2\\7913.4\\7859.5\\7846.6\end{array}\right)$&$\left(\begin{array}{c}8029.3\\7942.5\\7888.5\\7875.7\end{array}\right)$\\
$\frac12^{-}$ &$\left(\begin{array}{ccccc}-25.7&6.0&-20.0&4.1&-20.2\\6.0&46.2&20.8&-80.3&-14.3\\-20.0&20.8&72.2&-14.3&-74.4\\4.1&-80.3&-14.3&48.9&24.8\\-20.2&-14.3&-74.4&24.8&80.2\end{array}\right)$&$\left(\begin{array}{c}178.2\\100.4\\-50.1\\-26.5\\19.8\end{array}\right)$&$\left(\begin{array}{c}8219.9\\8142.1\\7991.6\\8015.2\\8061.5\end{array}\right)$&$\left(\begin{array}{c}7778.7\\7700.9\\7550.4\\7574.0\\7620.3\end{array}\right)$&$\left(\begin{array}{c}8029.4\\7951.6\\7801.1\\7824.7\\7871.0\end{array}\right)$&$\left(\begin{array}{c}8058.5\\7980.6\\7830.1\\7853.8\\7900.0\end{array}\right)$\\
\hline\hline\multicolumn{7}{c}{$nnsc\bar{b}$ ($I=0$)} \\\hline\hline
$J^{P}$ & $\langle H_{CM} \rangle$ &Eigenvalue &Mass&$(B_c\Lambda)$
&$(\Lambda_cB_s)$&$(\Xi_cB)$\\\hline
$\frac52^{-}$ &53.2&53.2&8094.9&7644.0&7890.5&7933.5\\
$\frac32^{-}$ &$\left(\begin{array}{cccc}-11.1&17.6&-3.6&6.6\\17.6&23.4&-16.9&14.3\\-3.6&-16.9&-94.9&-71.5\\6.6&14.3&-71.5&-132.3\end{array}\right)$&$\left(\begin{array}{c}-187.5\\-46.8\\37.7\\-18.4\end{array}\right)$&$\left(\begin{array}{c}7854.2\\7994.9\\8079.4\\8023.3\end{array}\right)$&$\left(\begin{array}{c}7403.3\\7544.0\\7628.5\\7572.4\end{array}\right)$&$\left(\begin{array}{c}7649.8\\7790.5\\7875.0\\7818.9\end{array}\right)$&$\left(\begin{array}{c}7692.7\\7833.5\\7918.0\\7861.9\end{array}\right)$\\
$\frac12^{-}$ &$\left(\begin{array}{ccccc}-49.7&-2.3&24.0&4.1&-20.2\\-2.3&-138.3&-9.0&-80.3&-14.3\\24.0&-9.0&-100.6&-14.3&-74.4\\4.1&-80.3&-14.3&-183.7&-7.5\\-20.2&-14.3&-74.4&-7.5&-140.6\end{array}\right)$&$\left(\begin{array}{c}-253.2\\-189.2\\-84.4\\-71.0\\-15.1\end{array}\right)$&$\left(\begin{array}{c}7788.5\\7852.5\\7957.3\\7970.7\\8026.6\end{array}\right)$&$\left(\begin{array}{c}7337.6\\7401.6\\7506.4\\7519.8\\7575.7\end{array}\right)$&$\left(\begin{array}{c}7584.1\\7648.1\\7752.9\\7766.3\\7822.2\end{array}\right)$&$\left(\begin{array}{c}7627.1\\7691.0\\7795.9\\7809.3\\7865.1\end{array}\right)$\\
\hline
\end{tabular}
\end{table}

\begin{table}[!h]
\caption{Calculated CMI's and estimated pentaquark masses of the
$ssnc\bar{b}$ and $sssc\bar{b}$ systems in units of MeV. The masses
in the forth column are calculated with the effective quark masses
and are theoretical upper limits.}\label{mass-ssncbbar}
\begin{tabular}{c|cccccc}\hline
\multicolumn{7}{c}{$ssnc\bar{b}$ ($I=\frac12$)} \\\hline\hline
$J^{P}$ & $\langle H_{CM} \rangle$ &Eigenvalue &Mass&$(B_c\Xi)$
&$(\Omega_cB)$&$(\Xi_c B_s)$\\\hline
$\frac52^{-}$ &47.5&47.5&8267.8&7805.2&8089.5&8018.3\\
$\frac32^{-}$ &$\left(\begin{array}{cccc}-20.2&19.2&3.6&6.6\\19.2&15.7&16.3&16.2\\3.6&16.3&43.6&-71.5\\6.6&16.2&-71.5&33.1\end{array}\right)$&$\left(\begin{array}{c}110.0\\-42.4\\32.7\\-28.2\end{array}\right)$&$\left(\begin{array}{c}8330.3\\8177.9\\8253.0\\8192.1\end{array}\right)$&$\left(\begin{array}{c}7867.8\\7715.3\\7790.4\\7729.6\end{array}\right)$&$\left(\begin{array}{c}8152.1\\7999.7\\8074.8\\8013.9\end{array}\right)$&$\left(\begin{array}{c}8080.9\\7928.5\\8003.6\\7942.7\end{array}\right)$\\
$\frac12^{-}$ &$\left(\begin{array}{ccccc}-60.8&2.3&-23.1&4.1&-22.9\\2.3&7.8&24.9&-80.3&-16.2\\-23.1&24.9&40.5&-16.2&-74.4\\4.1&-80.3&-16.2&5.3&20.9\\-22.9&-16.2&-74.4&20.9&32.6\end{array}\right)$&$\left(\begin{array}{c}139.9\\-87.9\\-69.8\\58.0\\-14.9\end{array}\right)$&$\left(\begin{array}{c}8360.2\\8132.4\\8150.5\\8278.3\\8205.4\end{array}\right)$&$\left(\begin{array}{c}7897.6\\7669.8\\7687.9\\7815.7\\7742.9\end{array}\right)$&$\left(\begin{array}{c}8182.0\\7954.2\\7972.3\\8100.1\\8027.2\end{array}\right)$&$\left(\begin{array}{c}8110.8\\7883.0\\7901.0\\8028.8\\7956.0\end{array}\right)$\\
\hline\hline\multicolumn{7}{c}{$sssc\bar{b}$ ($I=0$)} \\\hline\hline
$J^{P}$ & $\langle H_{CM} \rangle$ &Eigenvalue
&Mass&$(B_c\Omega)$&$(\Omega_c B_s)$\\\hline
$\frac32^{-}$ &71.1&71.1&8470.0&8019.5&8203.8\\
$\frac12^{-}$ &$\left(\begin{array}{cc}46.1&41.0\\41.0&71.6\end{array}\right)$&$\left(\begin{array}{c}101.8\\15.9\end{array}\right)$&$\left(\begin{array}{c}8500.7\\8414.8\end{array}\right)$&$\left(\begin{array}{c}8050.2\\7964.3\end{array}\right)$&$\left(\begin{array}{c}8234.5\\8148.6\end{array}\right)$\\
\hline
\end{tabular}
\end{table}

\begin{figure}[!h]
\begin{tabular}{ccc}
\includegraphics[width=220pt]{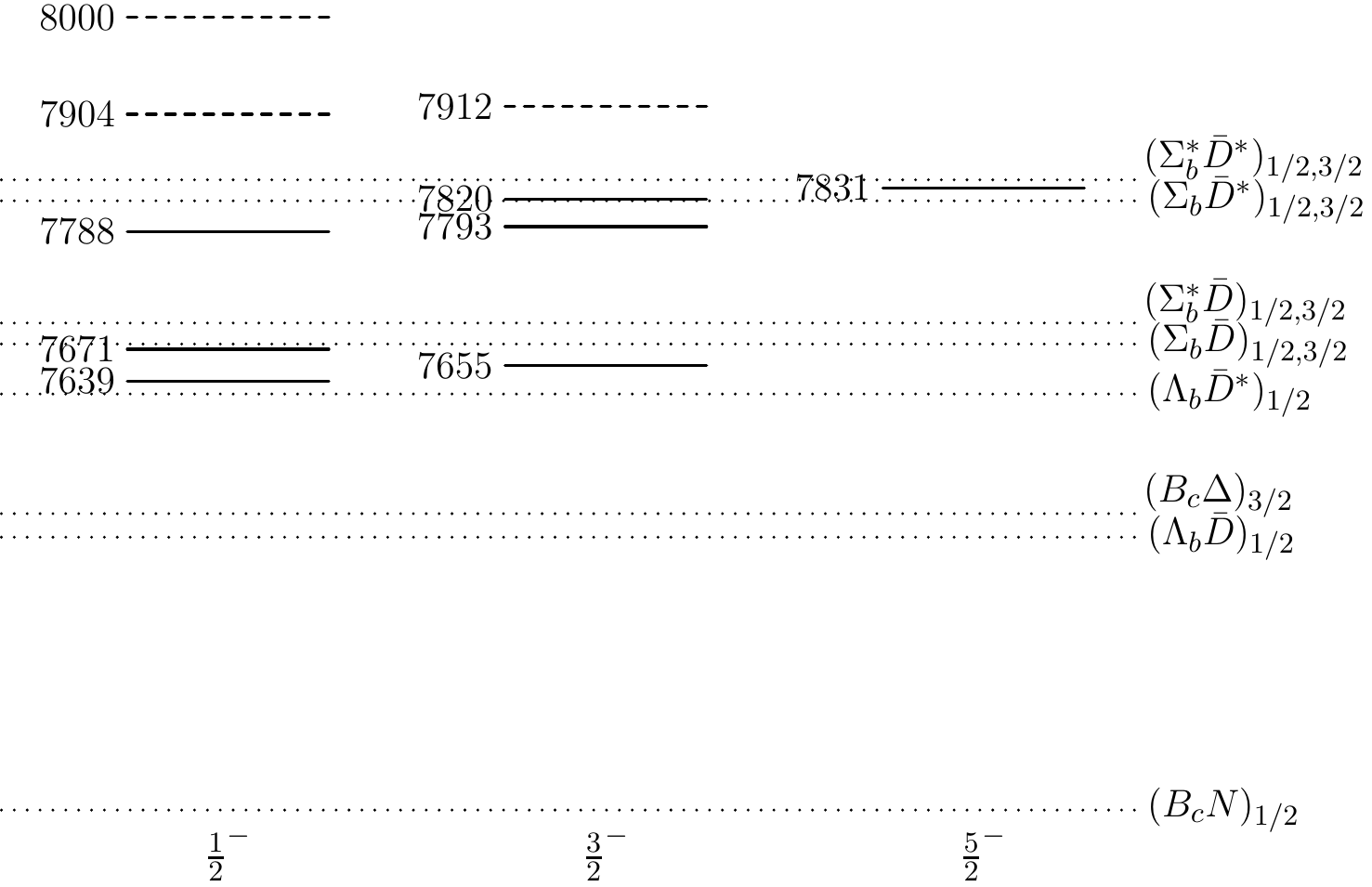}&\qquad&\includegraphics[width=220pt]{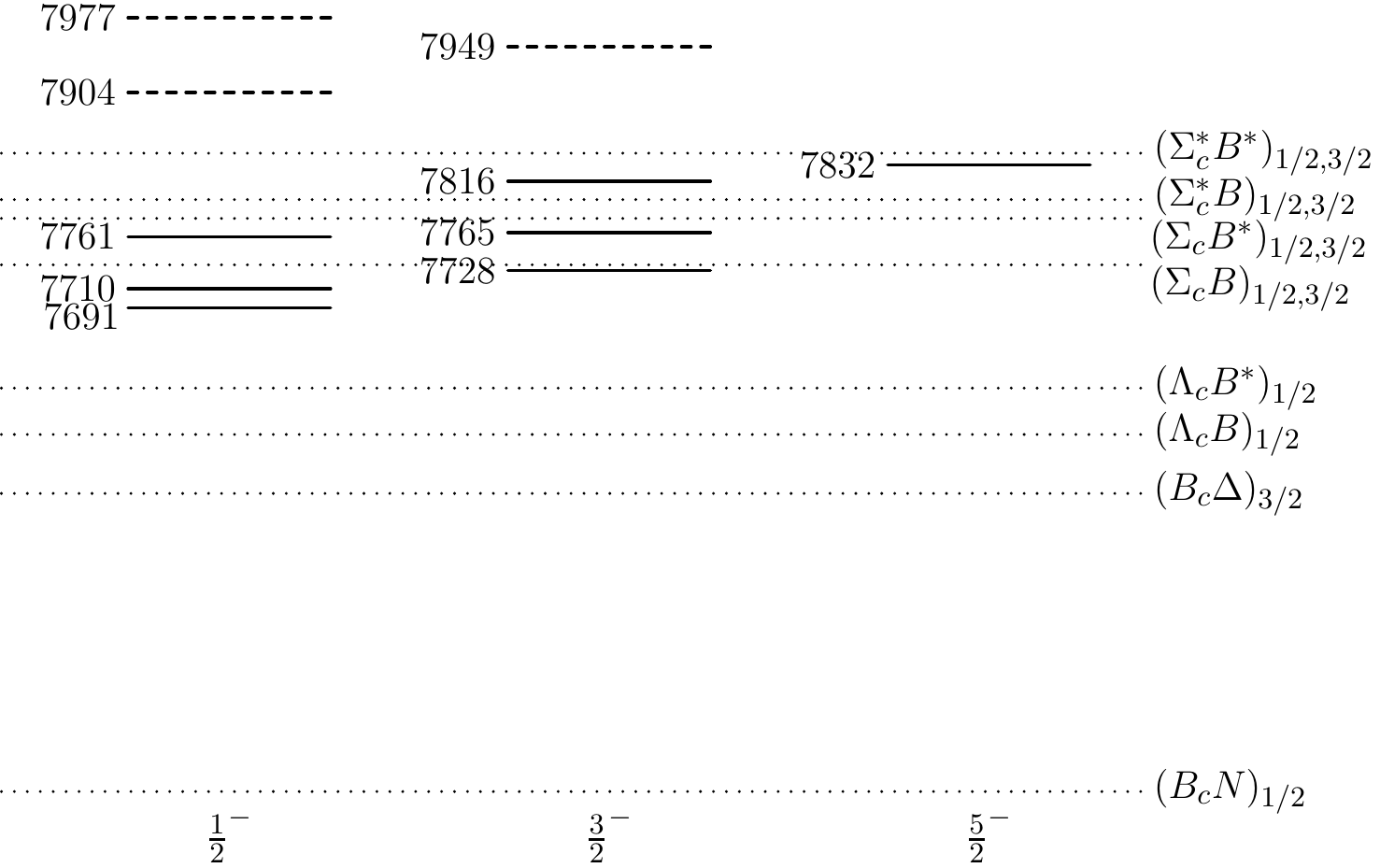}\\
(a) $I=\frac32$ (dashed) and $I=\frac12$ (solid) $nnnb\bar{c}$ states &&(b) $I=\frac32$ (dashed) and $I=\frac12$ (solid) $nnnc\bar{b}$ states\\&&\\
\includegraphics[width=220pt]{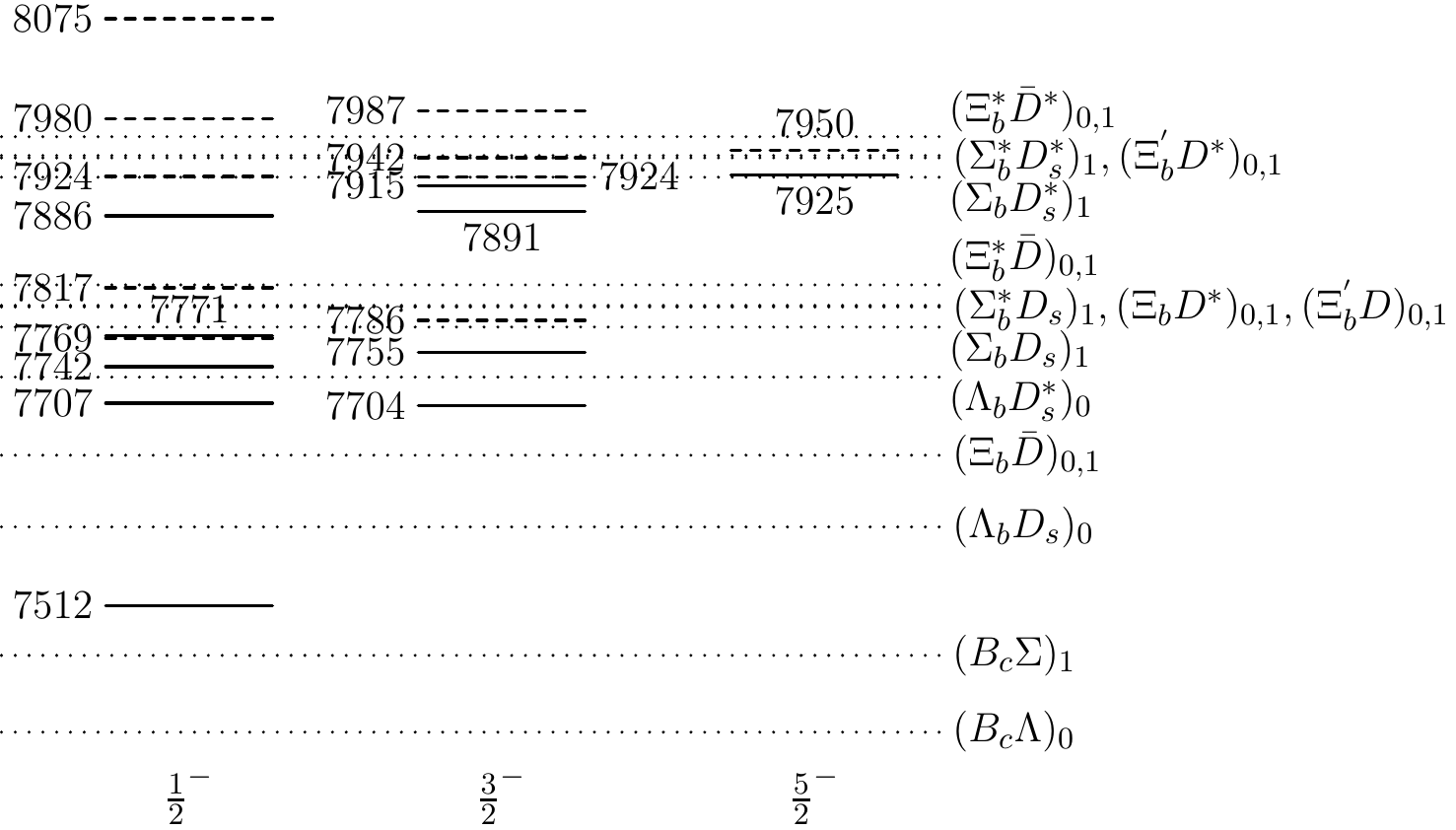}&\qquad&\includegraphics[width=220pt]{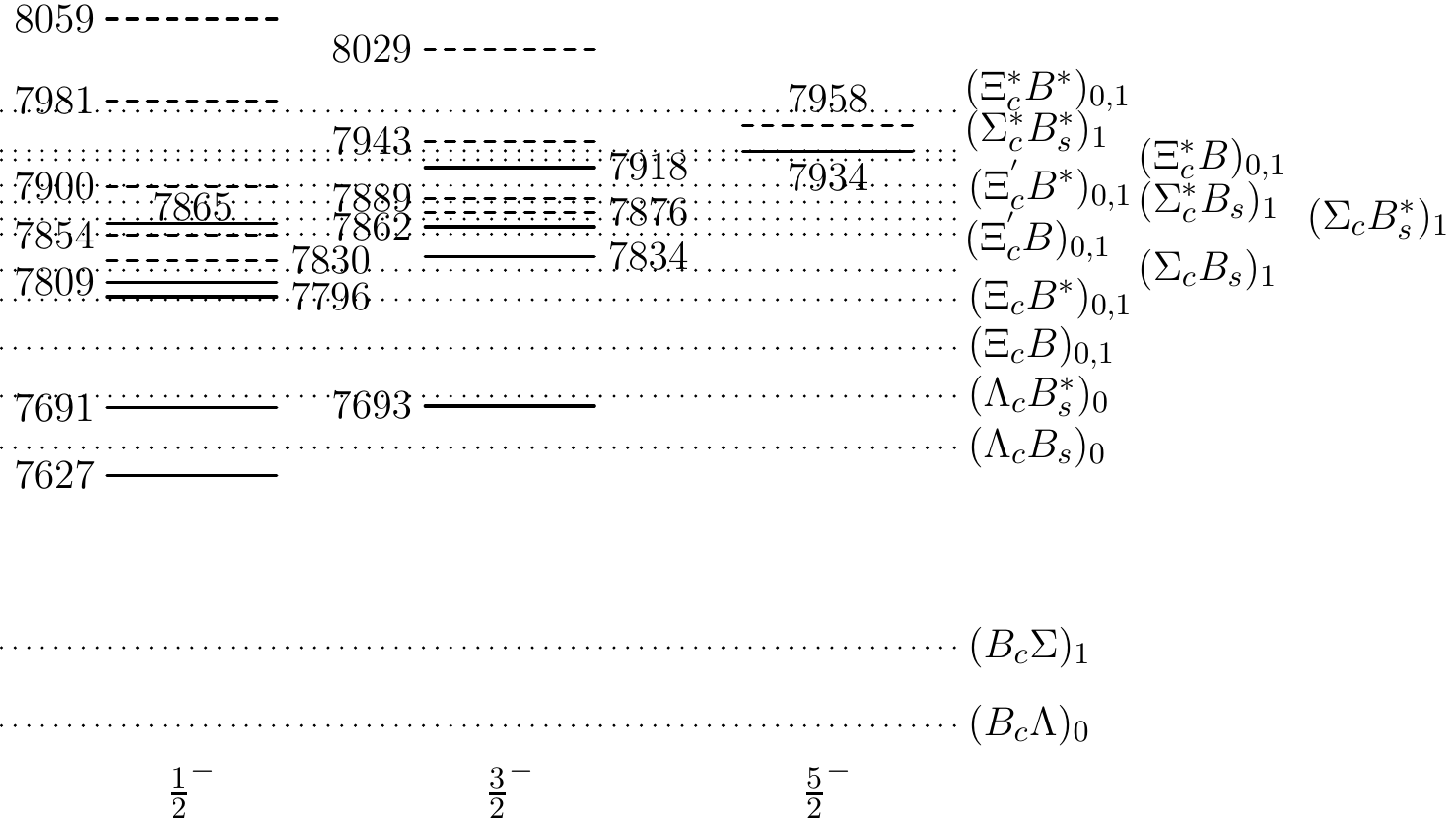}\\
(c) $I=1$ (dashed) and $I=0$ (solid) $nnsb\bar{c}$ states&&(d) $I=1$ (dashed) and $I=0$ (solid) $nnsc\bar{b}$ states\\&&\\
\includegraphics[width=220pt]{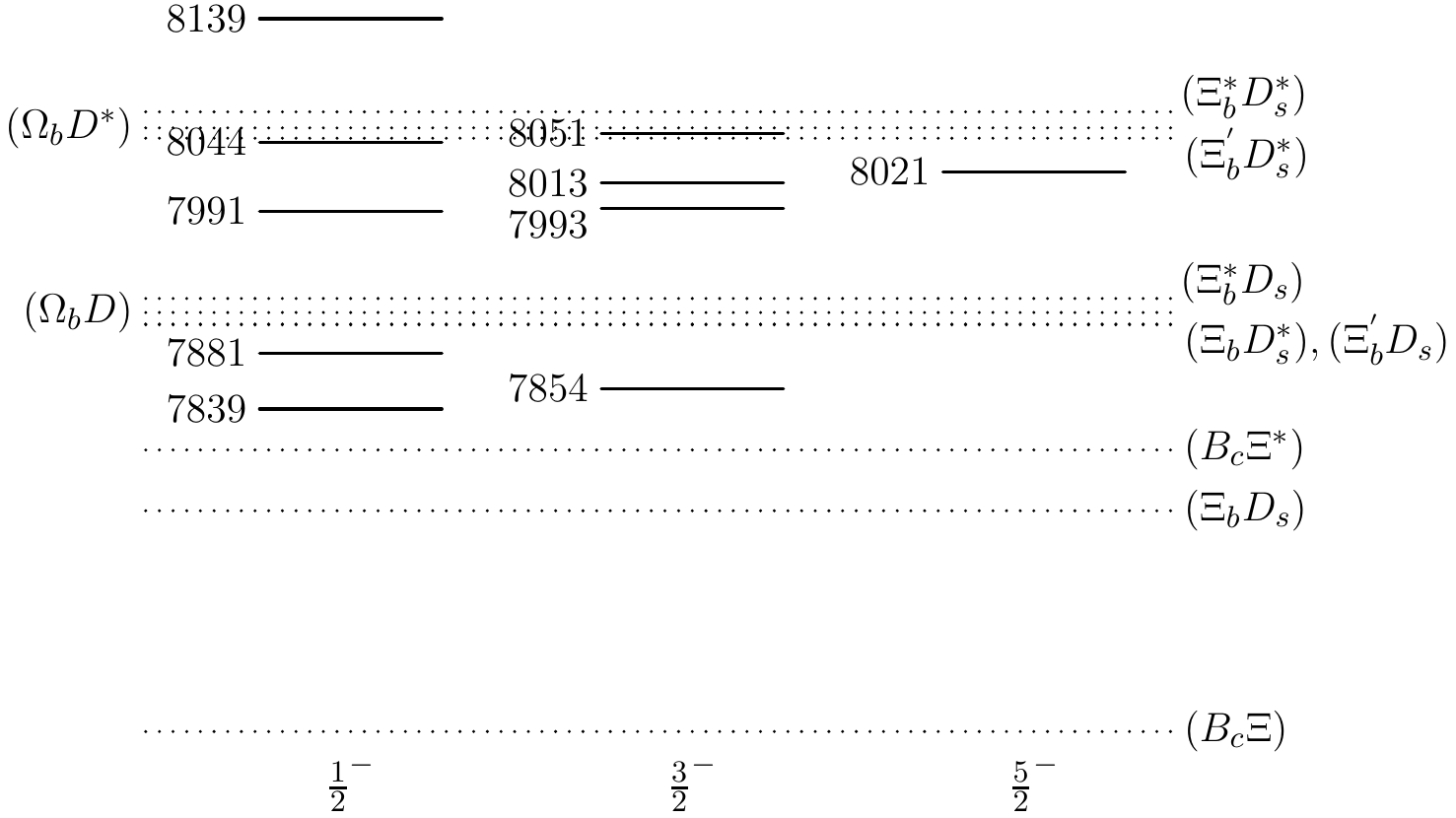}&\qquad&\includegraphics[width=220pt]{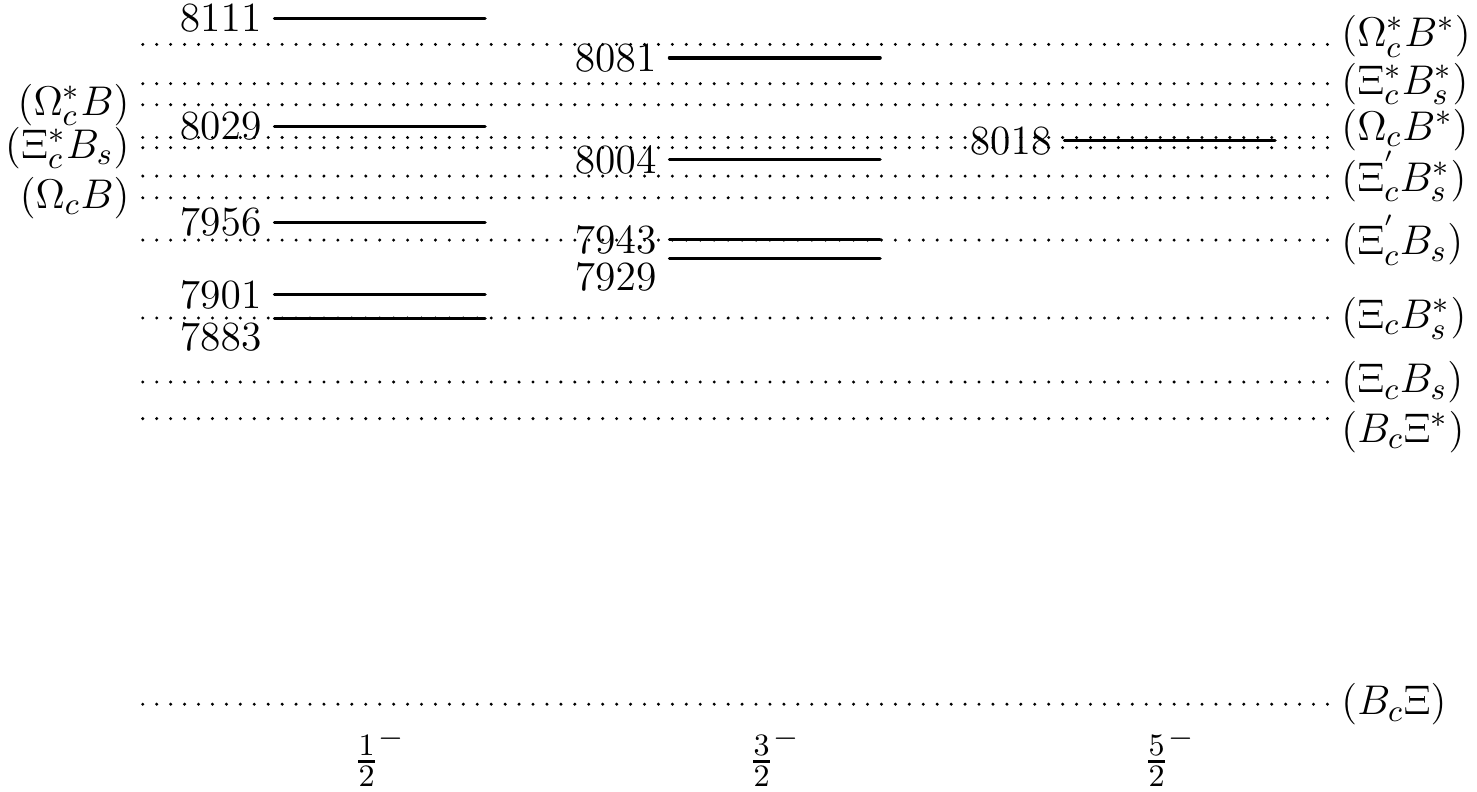}\\
(e) $I=\frac12$ (solid) $ssnb\bar{c}$ states&&(f) $I=\frac12$ (solid) $ssnc\bar{b}$ states\\&&\\
\includegraphics[width=220pt]{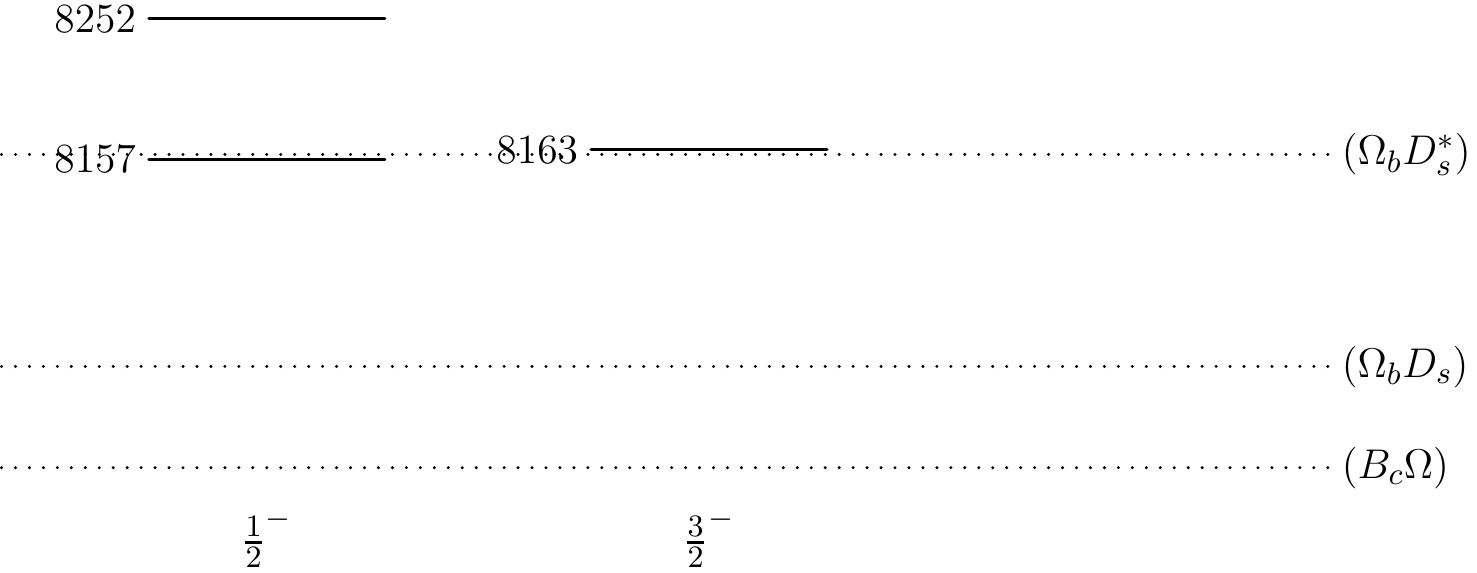}&\qquad&\includegraphics[width=220pt]{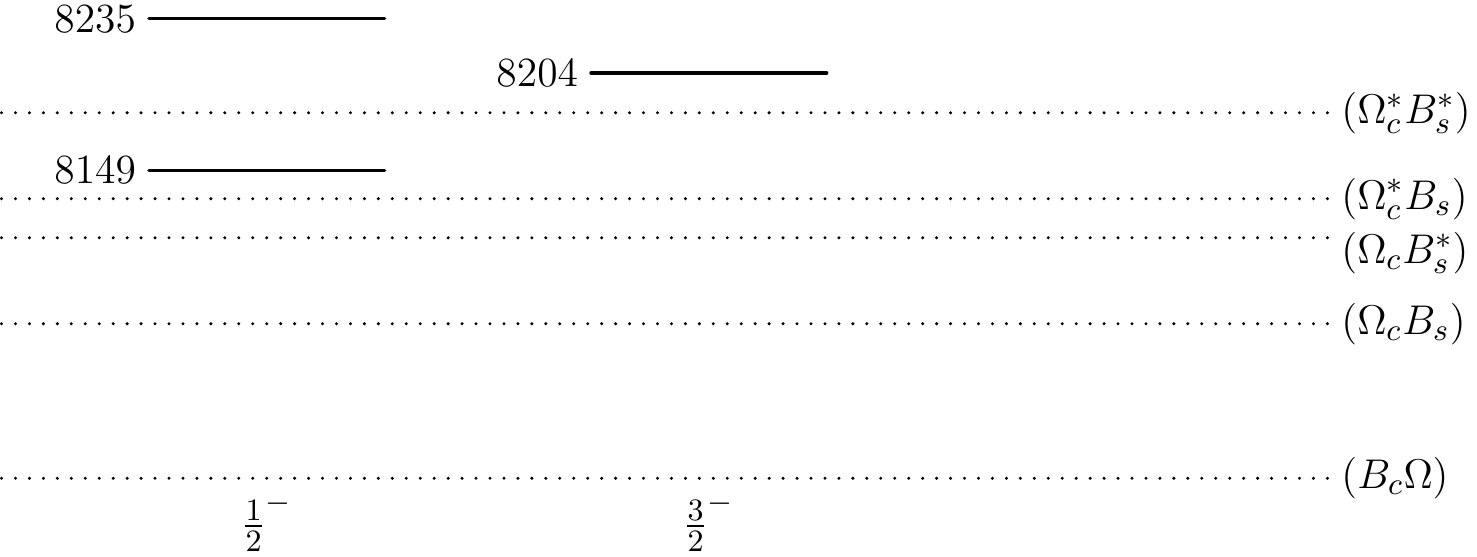}\\
(g) $I=0$ (solid) $sssb\bar{c}$ states&&(h) $I=0$ (solid)
$sssc\bar{b}$ states
\end{tabular}
\caption{Relative positions for the obtained $qqqb\bar{c}$ and
$qqqc\bar{b}$ pentaquark states. The dotted lines indicate various
meson-baryon thresholds. When a number in the subscript of a
meson-baryon state is equal to the isospin of an initial state, the
decay for the initial state into that meson-baryon channel through
S- or D-wave is allowed. We adopt the masses estimated with the
reference thresholds of $\Sigma_b\bar{D}$ (a), $\Sigma_cB$ (b),
$\Xi_b\bar{D}$ (c), $\Xi_cB$ (d), $\Xi_b D_s$ (e), $\Xi_cB_s$ (f),
$\Omega_bD_s$ (g), and $\Omega_cB_s$ (h). The masses are all in
units of MeV.}\label{fig-qqqbcbar}
\end{figure}

Now we move on to the more exotic $B_c$ like pentaquarks. In Ref.
\cite{Chen:2015loa}, possible $B_c$-like pentaquarks are explored.
Such baryons are lighter than the hidden-bottom states but heavier
than the hidden-charm partners. If their existence is possible, we
here give an estimation for the mass spectra of the $qqqb\bar{c}$
and $qqqc\bar{b}$ systems with colored $qqq$. These two types of
pentaquarks have slightly different masses. We present the numerical
results for the $nnnb\bar{c}$, $nnsb\bar{c}$, and $ssnb\bar{c}$
(also $sssb\bar{c}$) systems in Tabs. \ref{mass-nnnbcbar},
\ref{mass-nnsbcbar}, and \ref{mass-ssnbcbar}, respectively. Those
for the $nnnc\bar{b}$, $nnsc\bar{b}$, and $ssnc\bar{b}$ (also
$sssc\bar{b}$) systems in Tabs. \ref{mass-nnncbbar},
\ref{mass-nnscbbar}, and \ref{mass-ssncbbar}, respectively. For
comparison, we show the relative positions for these two types of
$B_c$-like pentaquarks in the same Fig. \ref{fig-qqqbcbar}. For the
two-body strong decays, all the channels involve flavored hadrons.
Here, when we say ``open-flavored channel'', for convenience, it
means that each final hadron contains a heavy (anti)quark.

For the $nnnb\bar{c}$ system, the lowest open-flavored decay channel
is $\Lambda_b\bar{D}$ ($\Sigma_b\bar{D}$) for the $I=\frac12$
($I=\frac32$) states. All the pentaquarks are above the thresholds
and are probably not narrow states. The decays of the
$J^P=\frac52^-$ state are again through $D$-wave. The $nnnc\bar{b}$
system has similar features. In Ref. \cite{Chen:2015loa}, the lowest
hadronic molecule is found to be $\Sigma_b\bar{D}$
($\Sigma_c\bar{B}^*$) in the case that the quark content is
$nnnb\bar{c}$ ($nnnc\bar{b}$). Then we here may conclude that more
pentaquarks below the molecules are possible. To distinguish a
molecule from a compact pentaquark, one needs both further
theoretical study and information from measured masses and decay
branching ratios. Experimentally, the search can be performed in the
$B_c^\pm N$ channels.

For the $nnsb\bar{c}$ and $nnsc\bar{b}$ systems, two $J^P=\frac12^-$
narrow states decaying into $B_c^\pm\Lambda$ are possible. If the
masses are overestimated, one finds more narrow states. To search
for them, the invariant mass distributions in the $B_c^\pm\Lambda$
channels should be studied. One may also search for relatively
narrow pentaquarks in the $B_c^\pm\Sigma$ channels.

For the $ssnb\bar{c}$ and $ssnc\bar{b}$ systems, all the
$J^P=\frac12^-$ and $\frac32^-$ pentaquarks seem to be broad states
unless the masses are overestimated. Those with $J^P=\frac52^-$ are
probably not-so-broad states. The search in the $B_c^\pm\Xi$
channels may provide useful information.

All the pentaquarks in the $sssb\bar{c}$ and $sssc\bar{b}$ systems
are probably broad states. It seems not easy to study them
experimentally.

In the hidden-charm and hidden-bottom cases, the colored heavy quark
pair may be generated from a gluon. However, in the $B_c$ case, the
two heavy quarks are produced from different gluons, which indicates
that the masses of the pentaquarks with colored $b\bar{c}$ or
$c\bar{b}$ are probably affected importantly by the colorless $B_c$
channels. The effects could be studied in a future work.

\section{Discussions and summary}\label{sec6}

Recently, the existence of heavy quark states with four quark
constituents has been confirmed. Heavy quark pentaquarks should also
exist from various theoretical calculations, where the less kinetic
energy is helpful to their formation. The observed hidden-charm
resonances $P_c(4380)$ and $P_c(4450)$ give an evidence for the
existence of pentaquarks, although the states still need further
confirmation. For a heavy quark many-body system, the number of the
spin partner states becomes large with the increasing number of
quarks and the mass splittings between these states should not be
large. The identification of such states will help us to understand
the formation of multiquark states in QCD. As a preliminary study,
we have here investigated the ground state compact pentaquarks
$qqqQ\bar{Q}$ ($q=u,d,s$ and $Q=b,c$) with colored $Q\bar{Q}$ in a
simple model.

We have constructed the flavor-color-spin wave functions for the
$nnnc\bar{c}$ ($n=u,d$) pentaquark states from the SU(3) symmetry
and calculated the color-spin interactions for the systems. Their
masses are estimated with different reference thresholds. Although
the study is not a dynamical calculation, the obtained mass spectrum
gives a basic feature for the possible pentaquarks. The constructed
wave functions are also helpful to study hadron decays in quark
models. After including the $SU(3)$ breaking effects and extending
the calculation to other heavy quark cases, we obtain systematic
results for various pentaquark systems.

There is a feature for the states we considered: the transition from
the colored $Q\bar{Q}$ into the colorless $Q\bar{Q}$ should be
suppressed and thus the branching ratios for the hidden-flavor
decays are small. Our results indicate that most pentaquarks have
open-flavor decay channels and should be broad states. The high spin
states with $J^P=\frac52^-$ probably have relatively narrow widths
because the $D$-wave decays are suppressed. In these pentaquark
states, an interesting observation is that the $nnsQ\bar{Q}$ systems
contain the low mass and thus narrow $\Lambda$-like states. To
search for them, the invariant mass distributions in the $J/\psi$
(or $\eta_c$, $\Upsilon$, $B_c$)+ $\Lambda$ channel are strongly
called for. In the literature, many investigations were performed in
the molecule models. Here we find that pentaquark states below the
molecules are also possible. Compared with possible baryon-meson
molecules, the compact pentaquarks with the colored $qqq$ cluster
should have smaller branching ratios for the hidden-flavor decay
channels.

In the present study, we have used the effective coupling constants
derived from the conventional hadrons. The couplings are related
with the hadron wave functions. But the wave functions in the
conventional hadrons and those in the multiquark states should be
different, one needs further investigations to answer whether this
extension is appropriate or not.

In short summary, we have estimated the masses of the hidden-charm
pentaquarks $qqqc\bar{c}$ with a color-magnetic interaction and
several reference thresholds, where the $qqq$ cluster is always a
color-octet state. Their hidden-bottom and $B_c$-like partners are
also investigated. In each case, we find that pentaquarks with lower
masses than hadronic molecules are possible. In the obtained baryon
states, the lowest ones seem to be the $J^P=\frac12^-$
$\Lambda$-like pentaquarks with suppressed decay channels, and the
$J^P=\frac52^-$ ones decaying through $D$-wave seem to be relatively
narrow, while more states have $S$-wave open-flavored decay channels
and should be broad. The masses and widths of the LHCb $P_c(4380)$
and $P_c(4450)$ are compatible with these features.

\section*{Acknowledgements}

YRL thanks Profs. M. Oka, S. Takeuchi, X.H. Zhong, Z.G. Si, and S.Y.
Li for helpful discussions. This project is supported by National
Natural Science Foundation of China under Grants No. 11175073. No.
11275115, No. 11222547, No. 11261130311 and 973 program. XL is also
supported by the National Program for Support of Young Top-notch
Professionals.

\end{document}